# CONVERSION OF TYCHO-2 TO JOHNSON-COUSINS MAGNITUDES IN THE GAIA ERA

STEPHEN C. SCHIFF
unaffiliated[1]

Subject: Instrumental and Methods for Astrophysics (astro-ph.IM)

## ABSTRACT

We take advantage of the availability of precision parallax data from Gaia Data Release 2 together with machine learning to develop a set of equations for transforming Tycho-2 (VT, BT) magnitudes into the Johnson-Cousins (J-C) system. Starting with data for 558 standard stars with apparent magnitudes brighter than 11.0, we employed one step supervised learning with weight decay regularization and 10-fold cross validation to produce a set of transformation equations from Tycho-2 into J-C, which in turn were used to derive transformations of the Tycho-2 standard deviations into the J-C system. Both the aggregated cross validation data sets and the in-sample results from the final training were essentially unbiased (average errors << 1 mmag in both B and V) and had error standard deviations comparable to those of the input data. Comparison of errors in- and out-of-sample indicate modest generalization error growth. Moreover, testing of the distributions of the normalized errors indicated that the predicted standard deviations are accurate, enabling them to be reliably employed in the suitability ranking of comparison star candidates. These results thus enable utilization of a substantial portion of the 2.5 million star Tycho-2 data set as comparison stars for two-color bright star ensemble photometry.

## 1. INTRODUCTION

This work was motivated by the need for accurate B and V band comparison stars in support of bright star (magnitude ≤ 11) ensemble photometry. In particular, the need was for a database of suitable comparison and check stars to enable formation of ensembles of circa 10 stars, matched to the target within 2 magnitudes, and bracketing it in color index, within a field of view that measures a few degrees on a side. The Tycho-2 database [ESO, 1997], [Høg, et al., 2000] provides the appropriate magnitude range and spatial density, but the Tycho-2 VT and BT filters are not precisely aligned with the corresponding V and B filters in the standard Johnson-Cousins (J-C) system. Consequently, we sought to find a method for converting the Tycho-2 magnitudes into their J-C counterparts in such a manner that we could also convert the Tycho-2 measurement standard deviations into corresponding J-C values. This latter requirement results from the need to choose the most accurate values from the list of candidate stars. Furthermore, we sought assurance that predictions based on one set of stars would accurately apply to others.

A number of Tycho-2 to J-C conversion algorithms have been published. [ESA, 1997]; [Bessell, 2000]; [Mamajek, Meyer & Leibert, 2002; 2006]. All of these are quite suitable for many purposes, but either lack universality (application across the Hertzsprung-Russell diagram), or the means for predicting estimated standard deviations or out-of-sample errors, and in some cases exhibit in-sample bias. (No criticism is intended: The tools and data that make this paper possible did not exist when they published their papers.) The present work is an attempt to remedy those limitations. In particular, our goals were to: (1) provide a tool that can be applied using readily available quantitative data; (2) allow estimation of error standard deviations based on those of the input data on a star by star basis, (3) include a means for estimating out-of-sample errors, and (4) reduce statistical bias.

### 1.1 Approach

Our approach to meeting the above stated goals involved two aspects that differentiate this work from that preceding it. We have abandoned reliance on spectral class and luminosity class taxonomies in favor of using parallax corrected magnitudes together with color indices. An underlying assumption, proven correct by the results, was that parallax corrected magnitudes are adequate surrogates for absolute magnitudes. Absolute magnitude together with color index $\chi \equiv$ BT-VT provides an unambiguous, quantitative mapping of the entire Hertzsprung-Russell diagram, and is therefore perfectly suited to the application of a quantitative mathematical approach. Implementation of our idea was made possible by the large scale, sub-milliarcsecond precision parallax measurement program carried out by the Gaia mission [Gaia Collaboration, 2016] and particularly by Gaia Data Release 2 [Gaia Collaboration, 2018].

The second distinctive aspect of our approach, as compared to earlier work, and enabled by the quantification of star characteristics, was to employ machine learning, in particular one step supervised learning. That method requires use of a data set comprised of known values to be fit by the hypotheses, said set ideally separated into a larger subset

---

[1] mailing address: PO Box 392, Aldie VA 20105; e-mail: astro_phys@runbox.com

# Conversion of Tycho-2 to Johnson-Cousins magnitudes in the Gaia era

employed to train the algorithm (the training set) and a smaller set used to measure its performance (the validation set). Validation set statistics provide a prediction of how well the final hypothesis will perform on data upon which it has not been trained (out-of-sample data). As will be discussed below, however, our data set was not of sufficient size to allow the ideal procedure to be followed, and so we employed a technique known as cross validation. A fundamental problem of fitting in general is that the data are noisy. In the present case, both the sample data (Tycho-2 VT and BT) and the standard stars (V and B) contain noise. To guard against fitting noise, we limit the number of degrees of freedom of our hypothesis set, and employ regularization to further smooth the fit.

In section (2), we discuss our methodology for training and validation data selection. In section (3) we discuss hypothesis selection and training methodology. Section (4) presents validation and full training results, including tabulation of the coefficients and equations for transforming Tycho-2 magnitudes and standard deviations into the J-C system. The appendix contains the complete data set used in performing the work.

## 1.2 Note on Terminology

Throughout this paper, we use the symbol $\chi$ to represent the quantity BT - VT, the color index in the Tycho-2 system. Magnitudes are represented in two ways: in upper case to represent apparent magnitudes, and in lower case to represent parallax corrected magnitudes. The letters M and m are used to designate apparent and parallax corrected magnitudes, respectively, when discussing general relationships. Stars having bt < 4.0 and $\chi$ > 0.6 are referred to as "GR" stars to avoid possible confusion with a conventional category that they encompass.

## 2. DATA SELECTION

We identified three sources of standard star data: [Menzies et al., 1989]; [Landolt, 1983]; and [Kilkenny et al., 1998]. The common underlying properties of these databases include numerous bright stars with measured V and B magnitudes in the J-C system with error standard deviations of less than 10 millimagnitudes (mmag), and typically with error standard deviations of 5 mmag.

The Tycho-2 data were selected from a single database, that of Høg, et al. [Høg, et al., 2000], downloaded from VizieR[2] Unless otherwise indicated, all data beyond those included in the three source papers were accessed via VizieR.

---
2  http://vizier.u-strasbg.fr/viz-bin/VizieR

All candidate data were subjected to a quality control procedure before being accepted. This procedure included culling out stars that were too dim (dimmer than magnitude 11.0) because the accuracy of the Tycho-2 data diminish with magnitude. While there are Tycho-2 stars dimmer than 11.0 that have suitable (< 50 mmag) standard deviations, we chose that limit so as to have a clearly delineated boundary on the applicability of our results. Then, known variable and double stars (including spectroscopic doubles) were eliminated. This involved, in the case of double stars, a star by star search through the Tycho Double Star catalog [Fabricius, et. al. 2002], and in case of variables, a star by star search through the General Catalog of Variable Stars [Samus', 2017][3]. A further culling was necessary in the case of the Kilkenny stars. Within that database, there were a number of stars having magnitudes based on very limited numbers of observations. Assuming the errors to be normally distributed, we performed a $\chi^2$ test [Sachs, 1984; p. 259] using sample size and sample standard deviation to test H0: the population standard deviation is less than or equal to 0.010 magnitudes at $\alpha$ = 0.05, retaining only those stars for which we failed to reject H0. At the end of the process we were left with 558 stars, the details for which are listed in the Appendix.

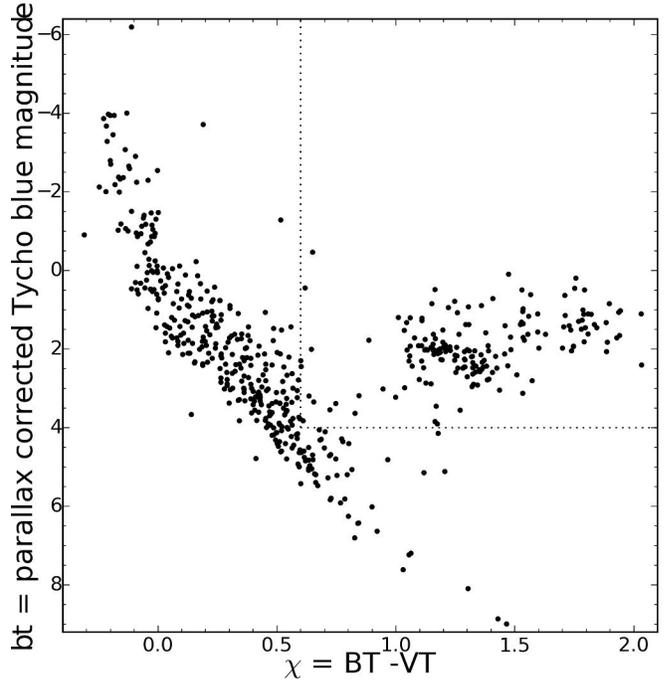

Figure (1): Plot of bt versus $\chi$ for the data set

---
3  Accessed via VizieR or directly at
   http://www.sai.msu.su/groups/cluster/gcvs/gcvs/

Figure (1) depicts the location of the data points in ($\chi$,bt) space. The dotted lines divide the space into two domains as discussed in Section 3.1.

## 3. HYPOTHESIS SELECTION AND TRAINING METHODOLOGY

The fields of data science and physics often invoke different approaches when it comes to fitting equations to data; in the former, preliminary data exploration is generally considered to be data snooping and is discouraged. On the other hand, when underlying physical processes are at least partially understood, the discarding of unnecessary degrees of freedom leads to a reduction in overfitting and possibly to better performance in terms of error minimization, both in and out-of-sample.

### 3.1 Preliminary Data Exploration

As our approach had the novel aspect of discarding especially luminosity class information in favor of parallax corrected magnitudes, it was thought prudent to make plots of J-C versus Tycho-2 magnitudes to inform the subsequent development of fitting polynomials. A plot of v versus vt yielded a very nearly linear relationship across the range from vt = +7.53 to vt = - 6.08. As the spread of the data was reasonably small (computed $R^2$ = 0.999) there was good reason to expect a successful fit using a single polynomial.

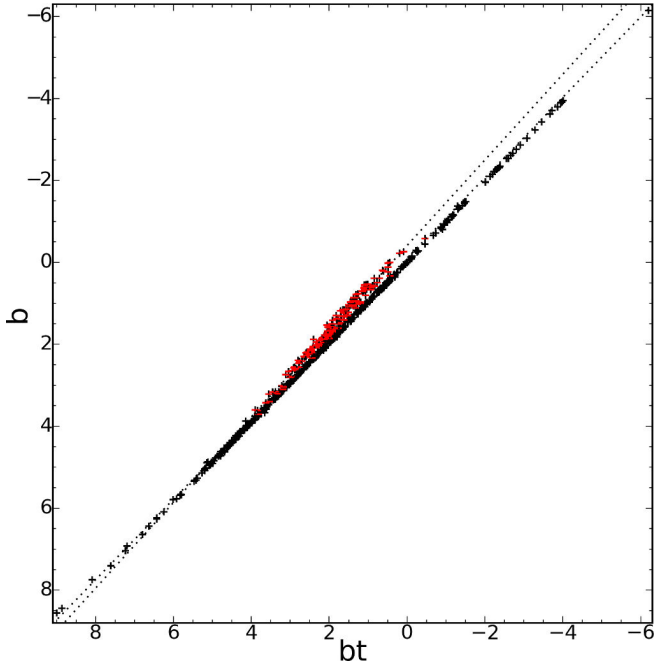

Figure (2): b versus bt for the data set

The plot of b versus bt, by contrast, yielded a configuration that would obviously require two polynomial fits. As indicated in Figure (2) the data, spanning a range in bt from +8.99 to -6.20, appear to be well fit ($R^2 \geq 0.99$) by a pair of lines. It was found by examination of the data that the line with greater slope (stars plotted in red) included stars having $\chi \geq 0.6$ and bt $\leq 4.0$, with some uncertainty due to the relative paucity of data points. Hereafter, we designate stars for which $\chi \geq 0.6$ and bt $\leq 4.0$ as GR[4] stars. The realm of the GR stars in ($\chi$, bt) space is depicted in Figure (1) as the region in the upper right bounded by the dotted lines.

### 3.2 Hypothesis Selection

From the point of view of data science, if the target function is to be fit by a polynomial of maximum degree n in two independent variables, then there are (n+1)(n+2)/2 degrees of freedom (polynomial terms) if the deleterious effects of data snooping are to be avoided. Add a degree of freedom for the single weight decay parameter $\lambda$ used in regularization, and the number of degrees of freedom becomes (n+1)(n+2)/2 + 1. Good data science practice suggests having at least 10 data points per degree of freedom [Abu-Mostafa, Magdon-Ismail and Lin, 2012, p.57]. Thus since the GR data set comprises but ca 180 stars, of which no more than ca 160 are available for training, we would be restricted to a 4th degree polynomial were we to follow the normal data science approach.

We argue that physics places natural constraints on the relationship between the Tycho-2 and J-C data that enable us to avoid the above conundrum. First, both measurement systems are linear, and for that reason we expect a strong linear relationship between the corresponding Tycho-2 and J-C magnitudes. This applies whether we form the relationships between apparent or parallax corrected magnitudes, because the parallax correction amounts to the same additive term applied to both quantities.

Second, we expect there to be a nonlinear relationship owing to the fact that the respective filter passbands are not identical, resulting in different responses even to a perfect blackbody, as a function of temperature and therefore of color index. When real world features such as absorption lines of varying quantity, location, width and depth are added to the mix, the relationship becomes not only more complex, but more sample-dependent. Thus we are motivated to use a hypothesis set in which there are linear terms in both parallax corrected magnitude and $\chi$, low order terms in their products, and higher order terms in $\chi$ only. While the earlier efforts cited above employed expansions to third degree in $\chi$, it is clear in retrospect that a higher order polynomial is necessary to reduce bias and increase scope. Ultimately, it was decided to employ a thirteen term polynomial including up to the 6th power of $\chi$. The particular polynomial chosen yields a data row vector of form

$$\mathbf{a} = [1,\ m,\ \chi,\ m^2,\ m\chi,\ \chi^2,\ m^3,\\ m^2\chi,\ m\chi^2,\ \chi^3,\ \chi^4,\ \chi^5,\ \chi^6] \quad (1)$$

where m = vt or bt according to which of the J-C magnitudes was being fit. The objective of the supervised learning exercise then became determining the values of the thirteen coefficients multiplying the terms in Equation (1) that minimize the out-of-sample error of fit. The parallax corrected, m values themselves were derived from the corresponding apparent magnitudes, M values using

$$m = M - 5\log_{10}\left(\frac{100}{plx}\right) \quad (2)$$

where plx is the Gaia DR2 [Gaia Collaboration, 2018] parallax in milliarcseconds, obtained via query of VizieR.

---
[4] GR is the abbreviation for "large reddish" in at least three languages.

### 3.3 Training, Regularization and Validation

To perform supervised learning so as to enable estimation of out-of-sample performance, the general procedure is to segment the available data into two sets, one of which is used to train, and the other to evaluate the results of training. In this context, the validation data are inaccessible to the training process, and therefore the results of training are diminished as a result of limitations to the size of the training set. In typical applications, this limitation is avoided by generating a larger data set. With a million samples, the harm done by holding back a few tens of thousands for validation is likely to be inconsequential. Unfortunately this is not the typical situation: our entire sample consisted of but N = 558 stars. To deal with such cases, cross validation is applied. According to that concept, the input data set is randomized then segmented into some number S equal parts. Training and validation are then performed S times, each time using a different one of the S data segments for validation and the rest for training. In this way, one obtains S independent validation sets, and consequently S independent estimates of out-of-sample error. The procedure is called S-fold cross validation, and S can vary from 2 to N-1. To provide meaningful statistics for a given validation set, S should be small, but to provide good training, S should be large. The common compromise choice of S in practice is 10 [Abu-Mostafa, Magdon-Ismail and Lin, 2012; p.150], and that is what we used: 10-fold cross validation. Our randomization process was to combine the three databases, then sort in order of increasing right ascension, on the theory that no independent variable is correlated with right ascension. To test the theory, we calculated the means and standard deviations for the two m variables and $\chi$ for each of the ten validation sets, and compared the results with the corresponding statistics of the data set as a whole. We failed to reject the hypothesis of equal means for 8 of the 10 sets of m values and all the $\chi$ mean, at $\alpha = 0.01$ using the t-test, and failed to reject the hypothesis of equal variances in all cases, using the F test, again at $\alpha = 0.01$. Thus the randomization process was judged adequate.

Both the Tycho-2 data and the standards to which we wish to fit them are noisy, and, being point samples, are not amenable to noise reduction through conventional filtering. Thus any attempt to fit a curve to the data risks fitting the noise, which in turn degrades the ability of the resultant fit to apply to out-of-sample data, for which the noise contributions are invariably different. This is referred to as overfitting. Choosing a hypothesis set that provides a large number of samples per degree of freedom is helpful but does not guarantee that overfitting will not occur. A more effective strategy is to employ regularization in addition to a large number of samples per degree of freedom. The idea is to smooth the curve of fit trading modest degradation in in-sample performance in the form of an augmented error for better out-of-sample performance (better generalization). In this work, we employed weight decay regularization.

Let $\mathbf{A}$ be the data matrix in which each row is a row vector of the form given by Equation (1). If $\mathbf{b}$ is the column vector containing the target (truth) values then the coefficient vector $\mathbf{x}$ that minimizes the error of fit in the least squares sense is the solution to

$$\mathbf{A}\mathbf{x} = \mathbf{b}\ .$$

As $\mathbf{A}$ is not square, this is solved by multiplying both sides on the left by the Moore-Penrose pseudoinverse:

$$(\mathbf{A}^T\mathbf{A})^{-1}\mathbf{A}^T\mathbf{A}\mathbf{x} = \mathbf{x} = (\mathbf{A}^T\mathbf{A})^{-1}\mathbf{A}^T\mathbf{b} \quad (3)$$

Equation (3) was employed to compute the coefficients for the non-regularized solutions in the 10-fold cross validation. In the above, error refers to the average value of the sum of the squared differences between the target values in $\mathbf{b}$ and the predicted values $\hat{\mathbf{b}}$ computed using the resulting coefficients vector $\mathbf{x}$:

$$\hat{\mathbf{b}} = \mathbf{A}\mathbf{x}$$

Schiff

In this work, our initial aim was to minimize two quantities: the average error and the standard deviation of the errors over the set. By computing the mean error and its standard deviation, the mean squared error comes for free, as  $\sigma_x^2 = \langle x^2 \rangle + \langle x \rangle^2$ .

For weight decay regularization, we utilize the fact that

$$x = (A^T A + \lambda I)^{-1} A^T b \qquad (4)$$

where **I** is the identity matrix, minimizes the augmented error in the least squares sense [Abu-Mostafa, Magdon-Ismail and Lin, 2012; p.133]. Repeated solution of Equation (4) for each training set, for differing values of λ, comprised the regularization process. It was conducted in two stages. In the first stage, trial and error selection of λ led to characterization of out-of-sample errors versus λ for a half dozen of the training sets. On that basis, we concluded that the range of λ that was likely to contain the values producing the minimum validation errors was 0.001 to 1.0. The second stage then consisted of training and validating all ten samples on a grid of λ values approximately evenly spaced in log(λ) over [0.001, 1.0]. This was in addition to training and validating for the non-regularized case λ = 0.

The regularization exercise produced optimum values of λ that varied not only over the validation sets but also according to the error criterion. Because the average errors were below the reporting sensitivity limit of 0.001 magnitude, the choice of λ (as a function of band) was based on minimizing the error standard deviation, which is essentially the same thing as minimizing the mean squared error. The three λ values thus determined were then employed in final training to compute the sets of coefficients for the three fits of Tycho-2 to J-C magnitudes, using the entire data set.

3.4 Transformation of Tycho-2- to J-C Standard Deviations

Any continuous function f(x,y, ...) can be expressed in terms of a Taylor series in its independent variables, which allows use of such expansions to express the statistics of a function of independent random variables for which their statistics are known. As it is generally sufficient to retain only the first order term [Bowker and Lieberman, 1972], we can calculate the standard deviations in the J-C system based on their known Tycho-2 counterparts using

$$\sigma_V^2 = \left[\frac{\partial V}{\partial v_T}\right]^2 \sigma_{VT}^2 + \left[\frac{\partial V}{\partial b_T}\right]^2 \sigma_{BT}^2 \qquad (5a)$$

and

$$\sigma_B^2 = \left[\frac{\partial B}{\partial v_T}\right]^2 \sigma_{VT}^2 + \left[\frac{\partial B}{\partial b_T}\right]^2 \sigma_{BT}^2 \qquad (5b)$$

Given that the noise processes involved in measuring the VT and BT magnitudes are independent, then [Feller, 1968]

$$\sigma_{B-V}^2 = \sigma_V^2 + \sigma_B^2 \qquad (6)$$

4. RESULTS

Here we summarize the results of the cross validation. Then, we summarize the results of applying the final hypothesis to the sample, indicating in-sample performance and estimate generalization error growth. Finally, we discuss how the actual errors in-sample corresponded to the standard deviations predicted by Equations (5).

4.1 Validation Errors

Table (1). Statistical summary of validation errors. Each table entry is calculated on the aggregate of the ten cross validation data sets, collectively spanning the entire data.

| Band | Average Error | Error Standard Deviation | Root Mean Squared Error |
|---|---|---|---|
| V | -4.84E-5 | 0.0215 | 0.0215 |
| B except GR | 2.02E-4 | 0.0181 | 0.0181 |
| B GR | 1.96E-4 | 0.0255 | 0.0255 |

Table (1) lists the results of error analysis on the aggregate of the ten validation sets for each of the three independent training exercises. That is to say, results from all ten validation sets were aggregated and statistics computed on the result, rather than computing statistics on each then averaging. All quantities are in magnitudes.

The validation errors observed are estimates of what the final training might be expected to achieve out-of-sample. In-sample performance was in all cases better, as would be expected, since the computation procedure minimizes in-sample mean squared error.

4.2 Final Hypothesis Results

The cross validation exercise had two objectives; to provide an estimate of out-of-sample error for the final hypothesis, and to provide a set of weight decay parameters to be used in training the final hypothesis. It was found that weight decay regularization improved the validation error performance as compared to not using weight decay regular-ization. Otherwise, the values of the weight decay parameters are of no importance to the results reported here.

The three best weight decay coefficients were employed to train the final hypotheses, applying Equation (4) with the entire 558 star data set. The coefficients corresponding to the **x** in Equation (1) are listed in Table (2).

Each data column in Table (2) corresponds to the coefficients to be applied to a row vector of the form given by Equation (1) to calculate the parallax corrected magnitude in the indicated band for the subject star. On a per-star basis, the vector represented by the appropriate column of Table (2) is dotted with that of the form of Equation (1) containing data for the star.

Table (2). Results of final training, in which Equation (4) was solved for the entire data set using the weight decay coefficients produced by cross validation regularization.

| Term: | V Coefficients | B, except GR Coefficients | B GR Coefficients |
|---|---|---|---|
| 1 | 1.0213E-2 | 4.0695E-4 | -4.9576E-2 |
| $m$ | 9.9424782E-1 | 9.9603370E-1 | 1.0189240E+0 |
| $\chi$ | -8.53859E-2 | -1.74110E-1 | -7.31677E-2 |
| $m^2$ | -1.09141E-3 | 1.43110E-3 | 1.42214E-3 |
| $m\chi$ | 2.5092800E-2 | -5.2396270E-3 | -4.9978470E-2 |
| $\chi^2$ | -1.42980E-1 | 6.12711E-2 | 1.78699E-2 |
| $m^3$ | 5.41208E-5 | 2.58949E-4 | -6.45412E-4 |
| $m^2\chi$ | -1.09966E-4 | -2.57778E-3 | 3.22588E-3 |
| $m\chi^2$ | -1.28451E-2 | -1.14915E-2 | 1.54743E-2 |
| $\chi^3$ | 1.200146E-1 | -3.210337E-3 | -5.003691E-2 |
| $\chi^4$ | 6.844508E-2 | 1.498395E-2 | -1.046353E-1 |
| $\chi^5$ | -9.432850E-2 | 1.193603E-2 | 8.671092E-2 |
| $\chi^6$ | 2.346223E-2 | -1.295700E-2 | -1.705617E-2 |

A note to programmers: In the process of exercising the final hypothesis against the data to compute the in-sample errors, two independently developed Python 3.6 with Numpy scripts were used. One of the scripts employed coefficients expressed as 64-bit floating point on a 64-bit machine; the other used coefficients expressed to lesser precision in the identical computing environment. The precision indicated in Table (2) was determined to be the minimum necessary to eliminate all discrepancies between the scripts' results in sample at the millimagnitude level.

Table (3) lists the in-sample performance of the final hypotheses when applied to the entire data set. It is seen that in all cases, in-sample performance is better than out-of-sample performance, as expected. However, the RMS error growth from in-sample to out-of-sample was only modest, amounting to no more than about 8%. Actual performance will vary according to sample size and composition, of course, and can be expected to sometimes be worse. It is seen that the average error grew by typically three orders of magnitude as a result of the weight decay regularization. This illustrates the bias-variance tradeoff.

Table (3). Statistical summary of in-sample errors, applying the coefficients of Table (2) to the entire data set.

| Band | Average Error | Error Standard Deviation | Root Mean Squared Error |
|---|---|---|---|
| V | 9.15E-8 | 0.0209 | 0.0209 |
| Other B | 7.58E-8 | 0.0173 | 0.0173 |
| B GR | -2.72E-7 | 0.0237 | 0.0237 |

Examination of Tables (1) and (3) reveals that both in terms of in- and out-of-sample performance, the final hypotheses are statistically unbiased. Particularly in the case of the more important validation errors, this was unexpected, though welcomed.

4.3 Reliability of the Standard Deviation Predictions

Taking the derivatives of Equation (1) and taking dot products with the appropriate column vectors in Table (3), then combining as indicated by equations (3) leads to estimates of standard deviations in the Johnson-Cousins system based on those in the Tycho-2 system. The question naturally arises as to the accuracy of those predictions. That question can be addressed, at least in part, by comparing the in-sample predictions and errors. For each sample, we divided the observed errors by the predicted standard deviations, for both the V and B bands. We expect that if the standard deviations are correct and the errors are normally distributed, the resulting ensemble will be normally distributed with mean 0 (because the errors are known to be unbiased) and standard deviation 1.

Testing took place in two steps. First, we tested the distribution of $\varepsilon_i/\sigma_i$ where $\varepsilon_i$ is the error of the ith sample and $\sigma_i$ is the predicted standard deviation for that sample, using the Kolmogorov-Smirnov test as applied by Lillifors (Sachs, 1984; Lillifors, 1967, pp.330-331). It was determined that in both cases (V and B) the null hypothesis, that the distributions were normal with empirical mean and standard deviation, could not be rejected at $\alpha = 0.01$. Having established that the distributions were indeed likely to be Gaussian, we tested the hypotheses that the means were zero and that the variances were equal to unity, using Student's t-test [Sachs, 1984, p.255] and the $\chi^2$ test [Sachs, 1984, p. 259], [Pearson and Hartley, 1976], respectively, failing on all occasions to reject the null hypothesis at $\alpha = 0.01$. It is therefore concluded that the predicted standard deviations are indicative, and can be used with confidence.

It is also of interest to compare the predicted (J-C) standard deviations with those associated with the Tycho-2 data. The predicted standard deviations are produced using equations (5), with partial differentiation applied to Equation (1), dotted with the coefficient vectors formed using the data in the appropriate columns of Table (2), squaring, summing, and finally taking a square root. As Equation (1) is a 6th degree polynomial, a 5th degree dependence is implicit in the results. Simply put, the resulting qualitative relationships between the predictions and their Tycho-2 counterparts are not intuitively clear.

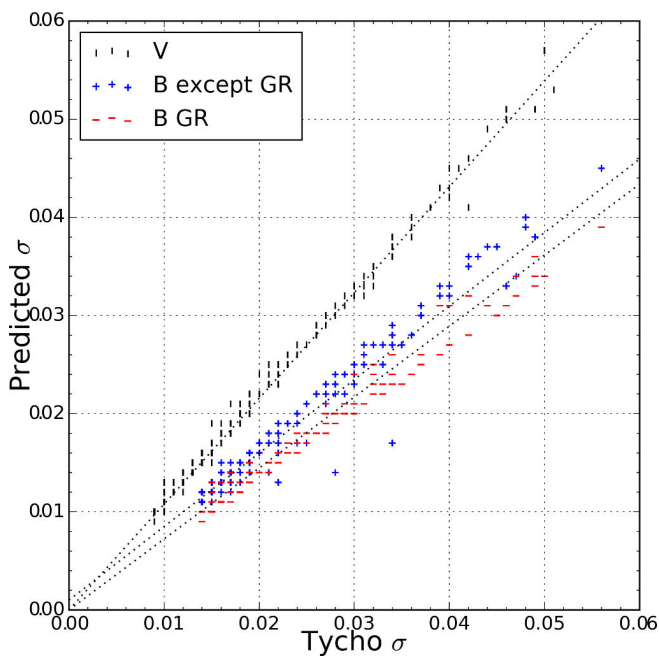

Figure (3): In-sample Relationship between Tycho-2 and J-C Standard Deviations

The actual in-sample relationship, as depicted in Figure (3), features two noteworthy elements. First, the relationship between $\sigma VT$ and $\sigma V$ is highly correlated and linear, with slope very close to unity. In-sample at least, we find that $\sigma VT$ is a good predictor for $\sigma V$. For the case of $\sigma B$ versus $\sigma BT$, on the other hand, the in-sample relationship is not nearly so well correlated, but noteworthy in that $\sigma BT$ is a pessimistic predictor for $\sigma B$, in-sample at least. The slopes of the two regression lines for blue stars are in the range 0.72 - 0.78, though the degree of pessimism is somewhat less in numerous instances.

### 4.4. Comments on Application

Although our sample was unrestricted in the sense that we did not exclude any particular region of the Hertzsprung-Russell diagram, it was of necessity a very small sample, and as a result some care needs to be taken in its use. Even a cursory examination of Figure (1) should be adequate to illustrate some major limitations, as there are regions that are either undersampled or not sampled at all. For example, there are very few stars in the training set with color indices greater than 1.8, discouraging application to very red stars. Quoting Bert Fowler, "The wear on a hypothesis is proportional to the degree of extrapolation." [Fowler, 1975].

Appendix A has been developed to aid in the process of applying our results, as well as to document the source data. To facilitate its use in determining whether a particular candidate comparison star is well embedded in the training set, the Appendix presents the data in order of increasing color index, rather than right ascension (as used in the cross validation process), and includes parallax-corrected Tycho magnitudes as opposed to apparent magnitudes.

### 4.5 Summary

The program described here has been successful in all respects. The polynomials are broadly applicable across a wide range of color indices and absolute magnitudes; data selection does not rely on knowledge of either spectral or luminosity class. Subsidiary equations, statistically proven in-sample, allow reliable estimation of the accuracy of the predictions. In-sample and validation set errors have been measured, and their statistics have been made public. Moreover, in-sample and validation testing demonstrated that the resulting equations are statistically unbiased, making the tools especially attractive for use in ensemble photometry.

### ACKNOWLEDGMENTS

The author would like to thank Professor S. George Djorgovski of Caltech for his role in the 2014 Caltech-JPL Summer School on Big Data[5] that introduced him to many new possibilities. Special thanks are due to fellow DSLR photometrist Mark Blackford, who graciously provided the Menzies, et al. data in machine readable form, and for his kind words of encouragement early in the effort.

---

5    https://archive.org/details/academictorrents_71268d279bcfcd9d88c8989c72158d8d73a2e2fc

## Appendix: The Database

For training, the data were sorted in order of increasing right ascension to randomize the magnitude and χ values to form training and validation sets. Here, the sort is by color index. In the table, the first column indicates the source; the next list the apparent magnitudes from the respective surveys ( [Kilkenny, et al., 1998]; [Landolt, 1983]; [Menzies, et al.,1989]). Right ascensions and declinations are either from the source or VizieR. Tycho-2 parallax corrected magnitudes, measured magnitude standard deviations, and color indices are listed next. The last column lists the Gaia DR-2 parallax in mas.

Table (A-1): Input data

| K, L, M | TYC | RA-2000 | Dec-2000 | V | B | bt | σBT | vt | σVT | χ | Plx |
|---|---|---|---|---|---|---|---|---|---|---|---|
| K | 7925 2575 1 | 19 07 07.89 | -41 43 16.0 | 10.190 | 9.986 | -0.912 | 0.028 | -0.601 | 0.042 | -0.311 | 0.6966 |
| K | 5808 665 1 | 22 05 31.16 | -13 46 12.8 | 9.842 | 9.647 | -2.135 | 0.028 | -1.887 | 0.040 | -0.248 | 0.4477 |
| K | 6440 462 1 | 02 53 40.80 | -26 09 20.4 | 8.494 | 8.256 | -3.866 | 0.016 | -3.636 | 0.013 | -0.230 | 0.3896 |
| M | 8166 2505 1 | 09 11 33.4 | -46 35 02 | 5.777 | 5.571 | -2.010 | 0.014 | -1.791 | 0.009 | -0.219 | 3.1233 |
| K | 8360 1779 1 | 18 35 15.91 | -45 56 27.0 | 8.376 | 8.155 | -3.676 | 0.015 | -3.459 | 0.013 | -0.217 | 0.4409 |
| K | 5822 692 1 | 23 21 50.85 | -09 45 41.0 | 7.681 | 7.467 | -3.286 | 0.016 | -3.071 | 0.012 | -0.215 | 0.7260 |
| K | 6524 2329 1 | 07 14 42.10 | -23 29 21.8 | 7.293 | 7.111 | -3.978 | 0.015 | -3.769 | 0.010 | -0.209 | 0.6247 |
| K | 8830 909 1 | 23 02 02.47 | -59 27 46.2 | 9.156 | 8.942 | -2.801 | 0.015 | -2.600 | 0.013 | -0.201 | 0.4579 |
| K | 6541 4018 1 | 07 19 09.91 | -25 33 57.2 | 7.463 | 7.284 | -3.948 | 0.014 | -3.747 | 0.010 | -0.201 | 0.5816 |
| K | 9311 86 1 | 20 19 06.34 | -69 26 06.4 | 8.846 | 8.658 | -2.711 | 0.016 | -2.512 | 0.012 | -0.199 | 0.5461 |
| M | 832 2490 1 | 14 58 31.9 | -43 08 02 | 2.669 | 2.443 | -3.456 | 0.014 | -3.267 | 0.009 | -0.189 | 6.7000 |
| K | 8262 2291 1 | 13 41 56.03 | -45 51 03.3 | 8.437 | 8.237 | -3.946 | 0.017 | -3.761 | 0.014 | -0.185 | 0.3733 |
| K | 7115 2422 1 | 07 14 46.90 | -37 14 43.4 | 7.224 | 7.054 | -2.192 | 0.015 | -2.010 | 0.010 | -0.182 | 1.4469 |
| K | 7055 995 1 | 05 32 27.80 | -30 33 16.2 | 9.438 | 9.274 | -1.031 | 0.017 | -0.863 | 0.018 | -0.168 | 0.8891 |
| M | 8121 1085 1 | 06 45 10.5 | -47 13 22 | 7.215 | 7.059 | -2.384 | 0.015 | -2.217 | 0.010 | -0.167 | 1.3209 |
| K | 8377 349 1 | 18 50 41.37 | -47 36 47.1 | 7.060 | 6.913 | -2.005 | 0.015 | -1.842 | 0.010 | -0.163 | 1.6774 |
| M | 7626 329 1 | 06 37 20.0 | -43 27 05 | 6.861 | 6.711 | -2.330 | 0.014 | -2.171 | 0.010 | -0.159 | 1.5802 |
| K | 8360 1598 1 | 18 39 42.41 | -45 14 09.3 | 9.403 | 9.259 | -1.190 | 0.019 | -1.033 | 0.022 | -0.157 | 0.8297 |
| K | 7972 1099 1 | 20 55 02.30 | -41 24 39.4 | 9.557 | 9.405 | -2.371 | 0.019 | -2.225 | 0.021 | -0.146 | 0.4540 |
| K | 8342 2924 1 | 17 34 04.19 | -46 36 01.7 | 7.657 | 7.534 | -3.079 | 0.016 | -2.940 | 0.012 | -0.139 | 0.7703 |
| M | 8108 1789 1 | 06 40 47.5 | -47 37 17 | 6.652 | 6.514 | -1.072 | 0.014 | -0.936 | 0.010 | -0.136 | 3.0974 |
| K | 8957 3002 1 | 10 34 49.51 | -60 11 14.0 | 8.542 | 8.458 | -4.007 | 0.016 | -3.876 | 0.013 | -0.131 | 0.3297 |
| L | 5819 618 | 22 50 28.24 | -13 18 44.6 | 10.112 | 9.997 | -1.015 | 0.025 | -0.888 | 0.034 | -0.127 | 0.6296 |
| K | 8707 223 1 | 15 20 32.52 | -59 32 37.6 | 7.322 | 7.237 | -2.658 | 0.015 | -2.534 | 0.010 | -0.124 | 1.0732 |
| K | 7062 1096 1 | 05 51 54.75 | -33 14 43.8 | 9.943 | 9.779 | -2.597 | 0.020 | -2.477 | 0.024 | -0.120 | 0.3444 |
| M | 8282 2757 1 | 14 28 51.9 | -47 59 32 | 6.396 | 6.286 | 0.463 | 0.014 | 0.575 | 0.010 | -0.112 | 6.9252 |
| K | 8958 2351 1 | 10 56 35.79 | -61 43 32.3 | 7.149 | 7.068 | -6.195 | 0.015 | -6.084 | 0.010 | -0.111 | 0.2273 |
| L | 5192 2044 1 | 21 02 59.62 | -00 55 29.1 | 6.498 | 6.399 | -1.515 | 0.014 | -1.404 | 0.010 | -0.111 | 2.6611 |
| M | 7639 2313 1 | 06 48 43.1 | -43 48 05 | 7.341 | 7.230 | 0.300 | 0.015 | 0.396 | 0.010 | -0.096 | 4.1348 |
| K | 7360 212 1 | 17 03 17.93 | -31 36 52.6 | 6.971 | 6.897 | -2.911 | 0.015 | -2.815 | 0.011 | -0.096 | 1.1150 |
| M | 8170 886 1 | 09 20 01.1 | -46 57 01 | 7.275 | 7.177 | -0.958 | 0.015 | -0.866 | 0.010 | -0.092 | 2.3928 |
| M | 8230 1494 1 | 12 17 47.5 | -45 23 34 | 7.956 | 7.868 | -2.246 | 0.014 | -2.156 | 0.010 | -0.090 | 0.9608 |
| M | 7626 1029 1 | 06 37 25.0 | -44 58 34 | 8.059 | 7.974 | 0.486 | 0.016 | 0.575 | 0.012 | -0.089 | 3.2129 |

Conversion of Tycho-2 to Johnson-Cousins magnitudes in the Gaia era

| K, L, M | TYC | RA-2000 | Dec-2000 | V | B | bt | σBT | vt | σVT | χ | Plx |
|---|---|---|---|---|---|---|---|---|---|---|---|
| L | 5183 2029 1 | 20 56 18.24 | -03 33 42.2 | 6.566 | 6.490 | -0.111 | 0.015 | -0.022 | 0.010 | -0.089 | 4.8463 |
| M | 8171 1472 1 | 09 23 39.8 | -46 54 32 | 6.215 | 6.122 | -0.866 | 0.014 | -0.779 | 0.010 | -0.087 | 4.0660 |
| M | 7540 798 1 | 01 17 00.8 | -42 31 58 | 7.857 | 7.771 | 0.586 | 0.015 | 0.670 | 0.010 | -0.084 | 3.6829 |
| M | 7895 864 1 | 17 26 53.6 | -44 42 08 | 10.009 | 10.057 | 0.318 | 0.025 | 0.391 | 0.032 | -0.073 | 1.1478 |
| M | 7895 5293 1 | 17 24 13.1 | -44 09 45 | 5.110 | 5.058 | -0.937 | 0.014 | -0.865 | 0.009 | -0.072 | 6.4195 |
| M | 8216 2578 1 | 11 59 10.7 | -45 49 56 | 6.346 | 6.276 | -1.134 | 0.014 | -1.066 | 0.009 | -0.068 | 3.3239 |
| M | 7946 1999 1 | 19 56 05.2 | -44 00 44 | 7.166 | 7.094 | -1.340 | 0.015 | -1.278 | 0.010 | -0.062 | 2.0655 |
| K | 7652 496 1 | 07 36 14.33 | -42 04 56.2 | 8.215 | 8.163 | -1.414 | 0.015 | -1.356 | 0.012 | -0.058 | 1.2348 |
| M | 8118 1887 1 | 06 58 00.7 | -46 05 45 | 6.829 | 6.764 | 0.442 | 0.015 | 0.497 | 0.010 | -0.055 | 5.4890 |
| M | 7892 7564 1 | 17 38 08.4 | -42 52 49 | 6.091 | 6.034 | -0.464 | 0.015 | -0.409 | 0.010 | -0.055 | 5.0637 |
| M | 7895 775 1 | 17 26 31.9 | -44 37 46 | 7.879 | 7.839 | -1.179 | 0.015 | -1.127 | 0.011 | -0.052 | 1.5856 |
| M | 8886 305 1 | 05 13 25.1 | -65 14 10 | 8.372 | 8.311 | 0.244 | 0.017 | 0.293 | 0.014 | -0.049 | 2.4305 |
| L | 4686 1086 1 | 01 57 56.14 | -02 05 57.7 | 8.874 | 8.792 | 0.047 | 0.018 | 0.093 | 0.016 | -0.046 | 1.7796 |
| M | 7895 505 1 | 17 29 52.2 | -44 44 30 | 7.549 | 7.493 | -0.676 | 0.016 | -0.633 | 0.012 | -0.043 | 2.3396 |
| K | 8150 591 1 | 08 36 02.24 | -46 30 05.9 | 7.263 | 7.252 | -2.296 | 0.015 | -2.254 | 0.010 | -0.042 | 1.2477 |
| M | 8000 1497 1 | 22 30 15.7 | -42 17 28 | 6.922 | 6.877 | 0.961 | 0.014 | 1.003 | 0.010 | -0.042 | 6.5875 |
| M | 8170 1150 1 | 09 13 34.5 | -47 20 18 | 5.910 | 5.864 | -0.290 | 0.014 | -0.251 | 0.009 | -0.039 | 5.9021 |
| L | 550 1098 1 | 21 35 55.81 | 05 28 35.2 | 8.301 | 8.247 | -0.128 | 0.017 | -0.090 | 0.014 | -0.038 | 2.1096 |
| M | 7882 352 1 | 17 18 47.8 | -44 07 47 | 5.759 | 5.711 | 0.466 | 0.014 | 0.499 | 0.009 | -0.033 | 8.9918 |
| M | 8342 6346 1 | 17 35 39.6 | -46 30 20 | 4.575 | 4.548 | -0.735 | 0.014 | -0.703 | 0.009 | -0.032 | 8.8552 |
| M | 7897 430 1 | 17 40 49.2 | -44 57 42 | 8.420 | 8.398 | -1.467 | 0.018 | -1.438 | 0.016 | -0.029 | 1.0758 |
| M | 8278 3080 1 | 14 35 10.7 | -46 27 43 | 6.890 | 6.873 | -0.874 | 0.014 | -0.846 | 0.010 | -0.028 | 2.8404 |
| M | 7767 1797 1 | 12 08 43.1 | -44 55 34 | 8.061 | 8.029 | 0.002 | 0.015 | 0.030 | 0.011 | -0.028 | 2.4930 |
| M | 8891 3507 1 | 05 41 42.8 | -67 24 10 | 7.054 | 7.033 | -1.162 | 0.015 | -1.137 | 0.010 | -0.025 | 2.3397 |
| M | 9161 1021 1 | 05 05 06.7 | -68 05 11 | 7.859 | 7.826 | 0.531 | 0.015 | 0.555 | 0.011 | -0.024 | 3.4822 |
| M | 8283 4014 1 | 14 37 54.9 | -48 01 25 | 6.654 | 6.628 | -1.055 | 0.014 | -1.032 | 0.009 | -0.023 | 2.9287 |
| M | 8279 840 1 | 14 47 55.5 | -46 09 22 | 8.089 | 8.063 | 0.042 | 0.015 | 0.062 | 0.011 | -0.020 | 2.5013 |
| M | 7959 1864 1 | 20 02 10.3 | -44 27 58 | 7.915 | 7.873 | 0.483 | 0.015 | 0.501 | 0.011 | -0.018 | 3.3189 |
| M | 8233 3212 1 | 12 08 14.7 | -48 41 33 | 5.337 | 5.319 | -0.869 | 0.014 | -0.853 | 0.009 | -0.016 | 5.8040 |
| M | 8347 1817 1 | 17 41 16.2 | -46 55 19 | 5.789 | 5.778 | -0.947 | 0.014 | -0.935 | 0.010 | -0.012 | 4.5427 |
| M | 7626 833 1 | 06 38 48.1 | -43 49 29 | 7.562 | 7.532 | -0.255 | 0.015 | -0.243 | 0.011 | -0.012 | 2.7759 |
| M | 9315 1893 1 | 20 00 35.6 | -72 54 38 | 3.949 | 3.922 | 1.445 | 0.014 | 1.455 | 0.009 | -0.010 | 31.8668 |
| M | 8279 1293 1 | 14 42 02.5 | -46 17 53 | 7.355 | 7.337 | -1.307 | 0.015 | -1.298 | 0.010 | -0.009 | 1.8740 |
| M | 8167 384 1 | 09 25 40.7 | -45 35 40 | 9.697 | 9.679 | 0.595 | 0.021 | 0.601 | 0.022 | -0.006 | 1.5495 |
| M | 7754 1736 1 | 11 51 13.1 | -43 55 59 | 6.599 | 6.585 | 1.082 | 0.015 | 1.088 | 0.010 | -0.006 | 7.9305 |
| M | 9166 165 1 | 05 29 08.9 | -70 49 32 | 8.072 | 8.047 | -0.122 | 0.016 | -0.117 | 0.012 | -0.005 | 2.3226 |
| M | 7896 1673 1 | 17 33 06.1 | -44 29 56 | 8.233 | 8.255 | 0.000 | 0.018 | 0.004 | 0.017 | -0.004 | 2.2300 |
| M | 7626 1386 1 | 06 36 28.9 | -44 39 04 | 8.912 | 8.871 | 0.050 | 0.017 | 0.053 | 0.015 | -0.003 | 1.7178 |
| M | 7896 1403 1 | 17 36 40.1 | -44 33 53 | 8.110 | 8.106 | -2.552 | 0.016 | -2.549 | 0.014 | -0.003 | 0.7444 |
| M | 7626 1443 1 | 06 37 42.6 | -44 29 44 | 7.689 | 7.675 | 0.736 | 0.015 | 0.736 | 0.011 | 0.000 | 4.0892 |
| M | 8118 1714 1 | 06 58 41.7 | -45 46 03 | 6.219 | 6.216 | 0.708 | 0.014 | 0.707 | 0.010 | 0.001 | 7.9590 |
| M | 896 640 1 | 17 34 29.7 | -43 09 36 | 8.837 | 8.851 | -1.483 | 0.018 | -1.484 | 0.018 | 0.001 | 0.8588 |
| M | 8279 2074 1 | 14 40 01.3 | -45 44 37 | 8.055 | 8.077 | 0.259 | 0.015 | 0.248 | 0.012 | 0.011 | 2.7279 |

| K, L, M | TYC | RA-2000 | Dec-2000 | V | B | bt | σBT | vt | σVT | χ | Plx |
|---|---|---|---|---|---|---|---|---|---|---|---|
| M | 8392 1174 1 | 20 13 34.8 | -45 35 26 | 7.788 | 7.773 | 0.443 | 0.015 | 0.424 | 0.011 | 0.019 | 3.3938 |
| M | 8229 2839 1 | 12 07 32.7 | -45 34 52 | 7.312 | 7.324 | -0.075 | 0.015 | -0.097 | 0.010 | 0.022 | 3.3091 |
| M | 7832 496 1 | 14 53 56.9 | -44 55 49 | 8.507 | 8.535 | 0.566 | 0.016 | 0.541 | 0.012 | 0.025 | 2.5483 |
| M | 8104 2168 1 | 06 38 41.4 | -46 10 30 | 8.375 | 8.389 | 1.375 | 0.016 | 1.347 | 0.013 | 0.028 | 3.8987 |
| M | 8104 820 1 | 06 43 14.1 | -45 14 31 | 10.655 | 10.706 | 1.811 | 0.039 | 1.781 | 0.049 | 0.030 | 1.6698 |
| M | 7703 1012 1 | 09 23 25.5 | -43 36 05 | 8.419 | 8.436 | 0.379 | 0.016 | 0.348 | 0.010 | 0.031 | 2.4567 |
| M | 9463 1179 1 | 19 49 10.2 | -76 54 31 | 7.138 | 7.149 | 1.592 | 0.015 | 1.561 | 0.010 | 0.031 | 7.6604 |
| M | 8391 394 1 | 20 00 53.4 | -45 41 58 | 8.744 | 8.774 | 1.415 | 0.017 | 1.378 | 0.015 | 0.037 | 3.3305 |
| M | 8171 201 1 | 09 23 50.8 | -47 03 34 | 8.191 | 8.195 | 0.084 | 0.016 | 0.046 | 0.012 | 0.038 | 2.3743 |
| M | 7703 1224 1 | 09 22 27.8 | -44 06 08 | 8.154 | 8.174 | 0.166 | 0.016 | 0.123 | 0.011 | 0.043 | 2.4868 |
| M | 8279 2020 1 | 14 41 15.6 | -45 42 19 | 10.004 | 10.135 | 2.094 | 0.027 | 2.051 | 0.031 | 0.043 | 2.4663 |
| M | 9459 1323 1 | 19 52 53.1 | -76 10 46 | 7.977 | 8.015 | 1.455 | 0.015 | 1.406 | 0.011 | 0.049 | 4.8442 |
| M | 9239 473 1 | 12 01 50.8 | -73 17 55 | 6.856 | 6.905 | 1.669 | 0.015 | 1.614 | 0.010 | 0.055 | 8.9045 |
| M | 7832 2493 1 | 14 56 24.8 | -44 42 15 | 6.754 | 6.804 | 1.194 | 0.015 | 1.137 | 0.010 | 0.057 | 7.5158 |
| M | 9239 1582 1 | 12 17 06.6 | -73 32 02 | 8.645 | 8.724 | -0.053 | 0.016 | -0.113 | 0.013 | 0.060 | 1.7542 |
| M | 9161 1172 1 | 05 00 28.1 | -69 21 01 | 8.750 | 8.792 | 1.708 | 0.018 | 1.647 | 0.015 | 0.061 | 3.8163 |
| M | 8010 1199 1 | 22 49 51.2 | -44 25 25 | 8.056 | 8.103 | 0.841 | 0.016 | 0.778 | 0.012 | 0.063 | 3.4967 |
| M | 8279 2 1 | 14 48 19.9 | -45 40 12 | 10.102 | 10.168 | 1.259 | 0.023 | 1.191 | 0.026 | 0.068 | 1.6531 |
| M | 8284 587 1 | 14 52 43.3 | -47 00 52 | 8.181 | 8.237 | 0.528 | 0.016 | 0.450 | 0.012 | 0.078 | 2.8439 |
| M | 7703 362 1 | 09 23 11.0 | -44 57 54 | 7.937 | 7.990 | 1.707 | 0.016 | 1.628 | 0.011 | 0.079 | 5.4714 |
| M | 7959 560 1 | 20 03 51.4 | -43 50 01 | 9.247 | 9.315 | 1.901 | 0.018 | 1.818 | 0.017 | 0.083 | 3.2593 |
| L | 4865 239 1 | 08 52 57.55 | -00 09 30.1 | 8.641 | 8.693 | 0.344 | 0.019 | 0.260 | 0.016 | 0.084 | 2.1056 |
| L | 4865 136 1 | 08 52 34.04 | -00 39 48.9 | 10.140 | 10.296 | 1.797 | 0.030 | 1.713 | 0.036 | 0.084 | 2.0192 |
| M | 8283 3821 1 | 14 46 29.0 | -47 26 28 | 5.734 | 5.802 | 0.950 | 0.014 | 0.861 | 0.009 | 0.089 | 10.6132 |
| M | 7626 1266 1 | 06 36 26.6 | -44 32 28 | 8.511 | 8.609 | -0.122 | 0.017 | -0.212 | 0.014 | 0.090 | 1.7688 |
| M | 8003 1374 1 | 22 30 42.9 | -44 05 37 | 6.937 | 7.006 | 1.416 | 0.015 | 1.324 | 0.010 | 0.092 | 7.5206 |
| M | 9411 1067 | 11 45 44.6 | -76 51 37 | 7.889 | 7.963 | 1.642 | 0.016 | 1.547 | 0.011 | 0.095 | 5.4102 |
| M | 8167 953 1 | 09 31 18.7 | -45 36 53 | 8.154 | 8.228 | 0.715 | 0.015 | 0.616 | 0.012 | 0.099 | 3.1007 |
| M | 8341 414 1 | 17 20 50.6 | -45 25 15 | 7.252 | 7.340 | 2.133 | 0.015 | 2.030 | 0.011 | 0.103 | 8.9484 |
| M | 9156 1126 1 | 04 00 43.7 | -71 10 00 | 6.596 | 6.676 | 1.589 | 0.014 | 1.483 | 0.010 | 0.106 | 9.4901 |
| M | 7703 1101 1 | 09 20 33.2 | -44 16 59 | 8.020 | 8.113 | 1.853 | 0.016 | 1.745 | 0.011 | 0.108 | 5.5223 |
| M | 7895 263 1 | 17 29 49.8 | -44 58 11 | 9.400 | 9.460 | 1.269 | 0.025 | 1.155 | 0.028 | 0.114 | 2.2366 |
| M | 9412 79 1 | 12 30 56.1 | -75 24 03 | 7.509 | 7.626 | 0.113 | 0.015 | -0.002 | 0.011 | 0.115 | 3.1339 |
| M | 8166 113 1 | 09 21 03.7 | -45 02 31 | 7.637 | 7.731 | 0.877 | 0.016 | 0.754 | 0.011 | 0.123 | 4.2016 |
| M | 7946 286 1 | 19 51 52.0 | -44 21 58 | 9.892 | 9.998 | 1.295 | 0.028 | 1.170 | 0.032 | 0.125 | 1.7528 |
| M | 8033 1148 1 | 01 27 35.3 | -46 09 06 | 7.704 | 7.791 | 1.770 | 0.015 | 1.643 | 0.011 | 0.127 | 6.1294 |
| M | 7882 355 1 | 17 17 31.3 | -44 46 40 | 8.344 | 8.474 | 2.099 | 0.017 | 1.972 | 0.015 | 0.127 | 5.1993 |
| L | 4914 649 1 | 10 55 16.97 | -01 25 28.8 | 9.380 | 9.440 | 0.624 | 0.020 | 0.496 | 0.019 | 0.128 | 1.6731 |
| M | 9463 1068 1 | 19 46 34.7 | -77 05 14 | 8.690 | 8.796 | 0.906 | 0.016 | 0.776 | 0.013 | 0.130 | 2.6219 |
| M | 7583 875 1 | 04 02 16.5 | -44 39 46 | 8.198 | 8.326 | 0.531 | 0.016 | 0.399 | 0.012 | 0.132 | 2.7429 |
| M | 7690 1718 1 | 09 18 35.0 | -44 36 56 | 8.348 | 8.438 | 2.096 | 0.015 | 1.963 | 0.012 | 0.133 | 5.2903 |
| K | 7567 1018 1 | 03 13 15.13 | -43 54 59.1 | 10.806 | 10.891 | 3.655 | 0.042 | 3.516 | 0.051 | 0.139 | 3.6005 |
| M | 8282 22 1 | 14 37 08.3 | -47 00 55 | 8.780 | 8.899 | 0.702 | 0.016 | 0.562 | 0.013 | 0.140 | 2.2554 |

Conversion of Tycho-2 to Johnson-Cousins magnitudes in the Gaia era

| K, L, M | TYC | RA-2000 | Dec-2000 | V | B | bt | σBT | vt | σVT | χ | Plx |
|---|---|---|---|---|---|---|---|---|---|---|---|
| M | 8033 906 1 | 01 20 00.8 | -45 27 32 | 9.657 | 9.743 | 1.439 | 0.020 | 1.296 | 0.019 | 0.143 | 2.1760 |
| M | 9141 7910 1 | 00 38 40.6 | -73 08 15 | 6.864 | 6.985 | 1.852 | 0.015 | 1.708 | 0.010 | 0.144 | 9.2468 |
| M | 8046 834 1 | 01 41 41.2 | -50 02 19 | 6.640 | 6.769 | 1.133 | 0.015 | 0.988 | 0.010 | 0.145 | 7.3583 |
| M | 8447 1616 1 | 22 59 14.8 | -45 11 21 | 7.292 | 7.433 | 2.345 | 0.015 | 2.192 | 0.011 | 0.153 | 9.4649 |
| M | 8103 1750 1 | 06 33 13.3 | -45 22 43 | 8.050 | 8.174 | 2.320 | 0.016 | 2.166 | 0.012 | 0.154 | 6.6497 |
| M | 7828 3521 1 | 14 50 58.7 | -42 49 21 | 6.832 | 6.966 | 1.174 | 0.015 | 1.020 | 0.011 | 0.154 | 6.8680 |
| M | 8342 669 1 | 17 26 15.8 | -45 10 28 | 10.548 | 10.717 | -0.231 | 0.042 | -0.390 | 0.050 | 0.159 | 0.6303 |
| M | 8230 434 1 | 12 11 51.3 | -45 06 55 | 8.557 | 8.704 | 0.130 | 0.016 | -0.036 | 0.011 | 0.166 | 1.9071 |
| M | 9162 552 1 | 05 17 23.0 | -68 28 19 | 9.155 | 9.299 | 1.138 | 0.020 | 0.969 | 0.018 | 0.169 | 2.2994 |
| M | 7832 1721 1 | 14 56 14.8 | -44 05 32 | 7.888 | 8.047 | 1.998 | 0.014 | 1.824 | 0.011 | 0.174 | 6.0721 |
| M | 8342 5977 1 | 17 37 12.0 | -45 54 00 | 8.576 | 8.738 | 0.329 | 0.019 | 0.151 | 0.016 | 0.178 | 2.0520 |
| M | 9318 305 1 | 19 49 23.2 | -73 31 23 | 7.341 | 7.488 | 0.715 | 0.015 | 0.535 | 0.010 | 0.180 | 4.3170 |
| M | 8167 1780 1 | 09 26 40.3 | -46 07 55 | 8.588 | 8.740 | 1.942 | 0.017 | 1.758 | 0.012 | 0.184 | 4.3030 |
| M | 9158 615 1 | 03 30 07.9 | -74 35 39 | 7.527 | 7.683 | 1.799 | 0.015 | 1.613 | 0.011 | 0.186 | 6.5340 |
| M | 8011 1348 1 | 22 55 56.3 | -42 33 09 | 7.258 | 7.416 | 2.405 | 0.015 | 2.218 | 0.011 | 0.187 | 9.7681 |
| M | 8067 2032 1 | 03 54 37.5 | -46 25 01 | 6.902 | 7.066 | 2.365 | 0.015 | 2.177 | 0.010 | 0.188 | 11.3189 |
| M | 8229 1784 1 | 12 03 56.4 | -45 45 04 | 10.614 | 10.775 | 1.078 | 0.031 | 0.890 | 0.034 | 0.188 | 1.1217 |
| M | 7639 818 1 | 06 43 33.6 | -44 03 18 | 8.269 | 8.433 | 1.686 | 0.015 | 1.497 | 0.012 | 0.189 | 4.3867 |
| M | 7882 17 1 | 17 17 05.5 | -44 46 43 | 6.458 | 6.714 | -3.724 | 0.014 | -3.913 | 0.010 | 0.189 | 0.8231 |
| M | 7639 2310 1 | 06 48 18.2 | -43 48 03 | 7.416 | 7.576 | 1.804 | 0.015 | 1.607 | 0.010 | 0.197 | 6.8478 |
| M | 8167 362 1 | 09 26 25.4 | -46 07 32 | 7.995 | 8.175 | 1.417 | 0.016 | 1.218 | 0.011 | 0.199 | 4.3848 |
| M | 8229 581 1 | 12 06 05.3 | -45 53 25 | 8.537 | 8.707 | 1.903 | 0.016 | 1.701 | 0.011 | 0.202 | 4.2387 |
| M | 7583 1591 1 | 04 02 48.8 | -44 17 53 | 8.811 | 8.933 | 1.527 | 0.018 | 1.322 | 0.010 | 0.205 | 3.2119 |
| M | 7576 164 1 | 03 49 49.3 | -42 43 39 | 7.348 | 7.527 | 1.045 | 0.015 | 0.839 | 0.010 | 0.206 | 4.9489 |
| M | 8216 1504 1 | 11 57 47.3 | -46 47 12 | 8.830 | 9.008 | 0.522 | 0.016 | 0.316 | 0.014 | 0.206 | 1.9662 |
| M | 8104 1056 1 | 06 38 29.4 | -45 22 26 | 8.168 | 8.352 | 2.326 | 0.016 | 2.118 | 0.012 | 0.208 | 6.0924 |
| M | 7583 721 1 | 04 03 45.2 | -44 43 59 | 7.646 | 7.820 | 2.095 | 0.015 | 1.883 | 0.011 | 0.212 | 7.0372 |
| M | 8167 501 1 | 09 23 30.0 | -46 16 14 | 9.158 | 9.339 | 1.627 | 0.018 | 1.409 | 0.016 | 0.218 | 2.8065 |
| M | 8033 900 1 | 01 22 10.4 | -46 13 18 | 8.444 | 8.650 | 0.785 | 0.016 | 0.560 | 0.012 | 0.225 | 2.6168 |
| M | 9160 1618 1 | 04 34 35.0 | -72 44 01 | 6.999 | 7.205 | 1.026 | 0.015 | 0.800 | 0.010 | 0.226 | 5.6937 |
| M | 9170 2132 1 | 05 22 25.3 | -71 30 23 | 7.434 | 7.627 | 1.280 | 0.015 | 1.053 | 0.011 | 0.227 | 5.2726 |
| M | 7959 1925 1 | 20 03 29.4 | -44 35 36 | 8.322 | 8.514 | 1.157 | 0.017 | 0.930 | 0.012 | 0.227 | 3.3047 |
| M | 9363 826 1 | 03 24 36.9 | -76 44 47 | 6.807 | 7.007 | 2.234 | 0.015 | 2.006 | 0.010 | 0.228 | 10.8856 |
| M | 9158 1157 1 | 03 28 16.0 | -73 32 59 | 8.176 | 8.366 | 0.742 | 0.016 | 0.512 | 0.012 | 0.230 | 2.9051 |
| M | 8071 1595 1 | 04 03 39.4 | -47 51 43 | 6.631 | 6.836 | 1.148 | 0.015 | 0.914 | 0.010 | 0.234 | 7.1527 |
| M | 8166 1322 1 | 09 22 04.0 | -45 38 34 | 9.602 | 9.802 | 0.422 | 0.022 | 0.185 | 0.021 | 0.237 | 1.3060 |
| L | 4800 923 1 | 06 52 01.88 | -00 27 21.4 | 10.537 | 10.740 | 1.967 | 0.043 | 1.727 | 0.046 | 0.240 | 1.7401 |
| M | 7960 886 1 | 20 14 59.7 | -43 40 42 | 8.160 | 8.370 | 1.448 | 0.016 | 1.206 | 0.010 | 0.242 | 3.9992 |
| L | 117 727 1 | 05 57 37.29 | 00 13 43.9 | 9.783 | 9.988 | 1.862 | 0.028 | 1.619 | 0.029 | 0.243 | 2.3650 |
| M | 9165 886 1 | 04 52 24.2 | -69 24 14 | 8.468 | 8.682 | 1.424 | 0.017 | 1.180 | 0.013 | 0.244 | 3.4737 |
| M | 7690 2567 1 | 09 18 53.8 | -44 02 25 | 8.859 | 9.073 | 1.363 | 0.017 | 1.117 | 0.014 | 0.246 | 2.7673 |
| M | 9363 659 1 | 03 55 10.8 | -75 09 50 | 9.043 | 9.270 | 1.440 | 0.017 | 1.189 | 0.015 | 0.251 | 2.6655 |
| M | 8068 89 1 | 04 05 34.1 | -45 40 00 | 7.566 | 7.800 | 0.755 | 0.015 | 0.497 | 0.010 | 0.258 | 3.8171 |

| K, L, M | TYC | RA-2000 | Dec-2000 | V | B | bt | σBT | vt | σVT | χ | Plx |
|---|---|---|---|---|---|---|---|---|---|---|---|
| M | 9314 780 1 | 19 49 25.3 | -72 30 12 | 5.389 | 5.625 | 2.494 | 0.014 | 2.234 | 0.009 | 0.260 | 23.0320 |
| M | 8233 3211 1 | 12 03 06.7 | -48 11 42 | 6.615 | 6.849 | 0.763 | 0.014 | 0.502 | 0.010 | 0.261 | 5.9471 |
| M | 9412 431 1 | 12 27 37.1 | -75 28 21 | 8.024 | 8.255 | 2.842 | 0.016 | 2.579 | 0.011 | 0.263 | 8.0891 |
| M | 7548 698 1 | 01 36 09.0 | -44 05 23 | 9.706 | 9.953 | 2.729 | 0.022 | 2.465 | 0.022 | 0.264 | 3.5286 |
| M | 7832 398 1 | 14 55 25.8 | -44 20 03 | 7.988 | 8.220 | 1.257 | 0.015 | 0.993 | 0.011 | 0.264 | 3.9641 |
| M | 9159 1440 1 | 04 05 44.3 | -73 16 51 | 8.967 | 9.206 | 2.174 | 0.019 | 1.910 | 0.015 | 0.264 | 3.8517 |
| M | 8280 1159 1 | 14 52 02.7 | -45 27 22 | 9.416 | 9.633 | 1.115 | 0.021 | 0.850 | 0.018 | 0.265 | 1.9175 |
| M | 7960 283 1 | 20 11 04.7 | -43 28 28 | 7.694 | 7.926 | 2.118 | 0.016 | 1.852 | 0.011 | 0.266 | 6.7545 |
| M | 7703 1032 1 | 09 27 45.1 | -43 38 16 | 8.849 | 9.077 | 2.218 | 0.016 | 1.951 | 0.014 | 0.267 | 4.1375 |
| M | 8171 1124 1 | 09 31 19.0 | -47 57 09 | 6.530 | 6.790 | 3.005 | 0.015 | 2.726 | 0.010 | 0.279 | 17.1588 |
| M | 9158 127 1 | 03 43 24.6 | -74 00 31 | 7.534 | 7.778 | 2.376 | 0.015 | 2.093 | 0.011 | 0.283 | 8.0721 |
| M | 9158 1418 1 | 03 21 23.0 | -74 35 39 | 8.040 | 8.298 | 1.961 | 0.016 | 1.677 | 0.011 | 0.284 | 5.2815 |
| M | 8229 1865 1 | 12 11 12.0 | -46 23 37 | 9.476 | 9.730 | 2.585 | 0.019 | 2.296 | 0.017 | 0.289 | 3.6598 |
| M | 7960 1039 1 | 20 10 50.2 | -43 36 11 | 7.827 | 8.083 | 2.771 | 0.016 | 2.481 | 0.012 | 0.290 | 8.4453 |
| M | 8104 1503 1 | 06 41 18.6 | -45 16 31 | 8.986 | 9.244 | 2.884 | 0.019 | 2.593 | 0.015 | 0.291 | 5.2256 |
| M | 8405 2457 1 | 20 26 04.8 | -46 39 36 | 6.727 | 7.001 | 2.917 | 0.015 | 2.626 | 0.010 | 0.291 | 14.9408 |
| M | 8279 549 1 | 14 47 44.4 | -45 16 16 | 10.532 | 10.801 | 1.824 | 0.034 | 1.530 | 0.036 | 0.294 | 1.5796 |
| M | 8167 488 1 | 09 28 23.0 | -45 45 52 | 9.540 | 9.854 | 3.373 | 0.023 | 3.074 | 0.022 | 0.299 | 4.8304 |
| M | 9165 609 1 | 05 01 34.8 | -69 29 46 | 8.131 | 8.389 | 1.712 | 0.016 | 1.412 | 0.013 | 0.300 | 4.4933 |
| M | 8067 746 1 | 03 51 41.8 | -47 19 01 | 6.703 | 6.981 | 0.888 | 0.015 | 0.586 | 0.010 | 0.302 | 5.9032 |
| L | 5081 1018 1 | 17 44 50.69 | -00 08 04.6 | 8.477 | 8.782 | 0.961 | 0.018 | 0.659 | 0.015 | 0.302 | 2.6866 |
| M | 8233 541 1 | 12 10 11.4 | -46 59 15 | 8.436 | 8.704 | 3.029 | 0.017 | 2.725 | 0.012 | 0.304 | 7.0857 |
| M | 8104 624 1 | 06 34 39.5 | -46 30 36 | 8.578 | 8.863 | 1.656 | 0.017 | 1.347 | 0.014 | 0.309 | 3.5306 |
| M | 8391 2553 1 | 20 00 48.3 | -45 06 47 | 5.795 | 6.086 | 2.126 | 0.014 | 1.815 | 0.009 | 0.311 | 15.8324 |
| M | 8010 772 1 | 22 40 51.8 | -44 30 15 | 8.891 | 9.193 | 2.608 | 0.019 | 2.297 | 0.015 | 0.311 | 4.6870 |
| M | 9158 246 1 | 03 41 11.1 | -73 55 46 | 8.399 | 8.704 | 2.989 | 0.017 | 2.675 | 0.013 | 0.314 | 7.0833 |
| M | 8104 885 1 | 06 41 15.6 | -45 06 47 | 8.656 | 8.922 | 1.618 | 0.017 | 1.297 | 0.014 | 0.321 | 3.3522 |
| M | 7583 1622 1 | 04 07 24.7 | -44 00 01 | 9.666 | 9.917 | 2.555 | 0.024 | 2.231 | 0.023 | 0.324 | 3.3127 |
| M | 7626 1045 1 | 06 35 10.8 | -44 57 30 | 10.010 | 10.315 | 2.385 | 0.032 | 2.059 | 0.032 | 0.326 | 2.5110 |
| M | 7754 784 1 | 11 55 46.1 | -44 30 36 | 8.685 | 9.022 | 1.747 | 0.017 | 1.413 | 0.013 | 0.334 | 3.4013 |
| M | 9158 85 1 | 03 37 41.0 | -74 49 21 | 6.883 | 7.201 | 3.298 | 0.015 | 2.962 | 0.010 | 0.336 | 16.1716 |
| M | 8167 834 1 | 09 31 13.0 | -46 00 27 | 7.980 | 8.275 | 1.087 | 0.016 | 0.750 | 0.011 | 0.337 | 3.5385 |
| M | 7548 908 1 | 01 38 30.8 | -42 55 40 | 6.665 | 6.995 | 2.382 | 0.015 | 2.041 | 0.010 | 0.341 | 11.6610 |
| M | 7956 1021 1 | 20 18 15.8 | -42 51 36 | 7.014 | 7.338 | 3.275 | 0.015 | 2.933 | 0.010 | 0.342 | 15.0311 |
| M | 8391 2155 1 | 19 59 31.6 | -46 11 50 | 10.318 | 10.743 | 3.821 | 0.039 | 3.479 | 0.040 | 0.342 | 4.0232 |
| M | 8342 4898 1 | 17 31 41.8 | -45 32 28 | 8.504 | 8.835 | 2.484 | 0.020 | 2.136 | 0.017 | 0.348 | 5.2301 |
| M | 7767 2026 1 | 12 09 33.0 | -44 52 57 | 9.284 | 9.620 | 1.728 | 0.019 | 1.378 | 0.015 | 0.350 | 2.6098 |
| M | 8886 1588 1 | 05 19 49.2 | -65 35 29 | 9.450 | 9.794 | 2.813 | 0.024 | 2.462 | 0.021 | 0.351 | 3.9367 |
| M | 8068 303 1 | 04 08 12.4 | -45 13 03 | 8.947 | 9.287 | 1.612 | 0.017 | 1.260 | 0.014 | 0.352 | 2.8717 |
| M | 7583 445 1 | 04 12 43.2 | -44 40 51 | 8.567 | 8.883 | 2.210 | 0.017 | 1.856 | 0.013 | 0.354 | 4.5128 |
| M | 9162 792 1 | 05 13 36.6 | -68 01 18 | 8.400 | 8.725 | 2.989 | 0.018 | 2.629 | 0.014 | 0.360 | 6.8841 |
| M | 8229 200 1 | 12 06 32.8 | -45 27 12 | 10.000 | 10.366 | 2.288 | 0.026 | 1.927 | 0.023 | 0.361 | 2.3558 |
| M | 7540 176 1 | 01 10 40.8 | -42 47 03 | 9.893 | 10.217 | 3.096 | 0.023 | 2.735 | 0.021 | 0.361 | 3.6990 |

Conversion of Tycho-2 to Johnson-Cousins magnitudes in the Gaia era

| K, L, M | TYC | RA-2000 | Dec-2000 | V | B | bt | σBT | vt | σVT | χ | Plx |
|---|---|---|---|---|---|---|---|---|---|---|---|
| M | 9411 1193 1 | 11 37 15.6 | -75 53 48 | 5.640 | 5.984 | 3.018 | 0.014 | 2.656 | 0.009 | 0.362 | 24.8604 |
| M | 9318 204 1 | 19 48 58.3 | -74 07 37 | 8.233 | 8.566 | 2.681 | 0.016 | 2.319 | 0.011 | 0.362 | 6.4715 |
| M | 7583 104 1 | 04 08 11.6 | -42 51 14 | 8.300 | 8.625 | 3.314 | 0.016 | 2.950 | 0.012 | 0.364 | 8.4828 |
| M | 8230 183 1 | 12 12 14.3 | -46 24 57 | 8.435 | 8.759 | 1.984 | 0.016 | 1.617 | 0.011 | 0.367 | 4.2800 |
| M | 9460 606 1 | 20 31 12.8 | -75 15 36 | 8.649 | 8.966 | 1.964 | 0.016 | 1.592 | 0.012 | 0.372 | 3.8207 |
| M | 7831 1133 1 | 14 47 21.1 | -44 27 11 | 9.374 | 9.744 | 2.212 | 0.029 | 1.839 | 0.027 | 0.373 | 3.0114 |
| M | 9138 1911 1 | 01 01 10.5 | -71 32 20 | 9.773 | 10.114 | 1.646 | 0.027 | 1.268 | 0.027 | 0.378 | 1.9396 |
| M | 9239 2103 1 | 12 17 00.4 | -74 08 37 | 7.834 | 8.184 | 1.425 | 0.015 | 1.046 | 0.011 | 0.379 | 4.2953 |
| M | 8447 288 1 | 22 57 58.3 | -45 09 30 | 7.717 | 8.079 | 2.719 | 0.016 | 2.340 | 0.011 | 0.379 | 8.2021 |
| M | 7895 381 1 | 17 27 58.6 | -44 43 16 | 8.647 | 8.973 | 1.784 | 0.017 | 1.398 | 0.014 | 0.386 | 3.4856 |
| M | 8033 311 1 | 01 21 32.2 | -45 08 15 | 7.455 | 7.817 | 3.255 | 0.014 | 2.866 | 0.010 | 0.389 | 11.8918 |
| M | 8447 1042 1 | 22 46 04.4 | -45 50 31 | 8.010 | 8.386 | 2.892 | 0.015 | 2.503 | 0.012 | 0.389 | 7.6957 |
| M | 9318 540 1 | 19 44 17.1 | -74 27 21 | 7.610 | 7.976 | 2.408 | 0.015 | 2.018 | 0.011 | 0.390 | 7.4452 |
| M | 8447 1060 1 | 22 55 30.2 | -46 19 42 | 7.530 | 7.904 | 2.230 | 0.016 | 1.840 | 0.011 | 0.390 | 7.1359 |
| M | 8003 348 1 | 22 39 24.1 | -44 44 46 | 8.928 | 9.268 | 2.989 | 0.019 | 2.597 | 0.015 | 0.392 | 5.3220 |
| L | 4985 159 1 | 14 44 14.08 | -00 37 06.6 | 9.484 | 9.864 | 3.618 | 0.034 | 3.226 | 0.031 | 0.392 | 5.5413 |
| M | 8446 1369 1 | 22 34 47.8 | -45 47 04 | 7.776 | 8.134 | 3.072 | 0.016 | 2.673 | 0.010 | 0.399 | 9.3817 |
| M | 7540 721 1 | 01 10 04.8 | -44 29 44 | 9.451 | 9.816 | 3.410 | 0.020 | 3.010 | 0.016 | 0.400 | 5.0947 |
| L | 4932 143 | 11 56 06.01 | -00 47 54.3 | 9.862 | 10.231 | 3.319 | 0.029 | 2.914 | 0.028 | 0.405 | 4.0349 |
| L | 4966 1304 1 | 13 37 30.34 | -00 13 17.3 | 8.760 | 9.102 | 3.553 | 0.021 | 3.147 | 0.017 | 0.406 | 7.4862 |
| M | 8341 277 1 | 17 24 43.1 | -45 00 30 | 6.656 | 7.038 | 3.210 | 0.015 | 2.800 | 0.010 | 0.410 | 16.5890 |
| M | 7942 379 1 | 19 54 58.7 | -42 38 43 | 10.130 | 10.557 | 4.783 | 0.040 | 4.372 | 0.038 | 0.411 | 6.7535 |
| M | 9159 1624 1 | 03 48 11.5 | -74 41 39 | 7.125 | 7.518 | 3.806 | 0.015 | 3.392 | 0.010 | 0.414 | 17.5271 |
| L | 4833 597 1 | 07 55 13.01 | -00 25 32.8 | 9.807 | 10.214 | 2.150 | 0.042 | 1.731 | 0.040 | 0.419 | 2.4450 |
| M | 7704 2247 1 | 09 30 29.4 | -44 52 41 | 8.489 | 8.874 | 3.112 | 0.017 | 2.692 | 0.010 | 0.420 | 6.8203 |
| M | 9158 908 1 | 03 18 57.7 | -73 59 10 | 7.237 | 7.638 | 3.088 | 0.015 | 2.665 | 0.010 | 0.423 | 11.9244 |
| M | 9159 271 1 | 03 46 36.9 | -74 30 29 | 8.014 | 8.407 | 3.599 | 0.015 | 3.176 | 0.012 | 0.423 | 10.5609 |
| M | 7767 1471 1 | 12 08 39.0 | -44 52 18 | 10.086 | 10.462 | 2.786 | 0.029 | 2.363 | 0.027 | 0.423 | 2.8388 |
| M | 7576 1320 1 | 03 51 03.1 | -44 21 54 | 8.008 | 8.396 | 1.918 | 0.015 | 1.493 | 0.012 | 0.425 | 4.8843 |
| M | 8216 786 1 | 11 50 25.8 | -46 28 08 | 7.664 | 8.054 | 3.250 | 0.015 | 2.824 | 0.010 | 0.426 | 10.5820 |
| M | 7547 707 1 | 01 31 29.6 | -44 39 22 | 7.864 | 8.260 | 3.252 | 0.016 | 2.825 | 0.011 | 0.427 | 9.6192 |
| M | 8220 2883 1 | 11 56 43.9 | -47 04 21 | 6.256 | 6.660 | 3.156 | 0.014 | 2.729 | 0.010 | 0.427 | 19.2813 |
| M | 9412 215 1 | 12 32 50.2 | -75 26 05 | 7.785 | 8.182 | 3.317 | 0.015 | 2.887 | 0.011 | 0.430 | 10.2735 |
| M | 8447 192 1 | 22 46 52.1 | -45 03 05 | 8.240 | 8.600 | 3.215 | 0.017 | 2.785 | 0.013 | 0.430 | 8.0282 |
| M | 7763 1047 1 | 12 03 39.6 | -42 26 03 | 5.148 | 5.560 | 3.413 | 0.014 | 2.979 | 0.009 | 0.434 | 36.0179 |
| M | 8104 1496 1 | 06 38 15.5 | -45 16 42 | 8.278 | 8.686 | 2.962 | 0.017 | 2.526 | 0.012 | 0.436 | 6.9845 |
| M | 7767 105 1 | 12 07 22.9 | -43 14 47 | 7.207 | 7.618 | 2.652 | 0.015 | 2.212 | 0.010 | 0.440 | 9.8234 |
| M | 8731 1065 1 | 17 12 09.2 | -43 14 21 | 3.326 | 3.734 | 2.119 | 0.014 | 1.671 | 0.009 | 0.448 | 45.9556 |
| M | 7636 389 1 | 06 52 39.6 | -42 30 16 | 6.519 | 6.931 | 1.059 | 0.015 | 0.609 | 0.010 | 0.450 | 6.4545 |
| M | 8346 2791 1 | 17 29 54.2 | -47 00 23 | 8.379 | 8.818 | 3.810 | 0.017 | 3.360 | 0.014 | 0.450 | 9.6316 |
| M | 8166 2365 1 | 09 12 13.8 | -45 50 54 | 6.661 | 7.091 | 2.524 | 0.015 | 2.072 | 0.010 | 0.452 | 11.8181 |
| M | 7959 1223 1 | 20 00 51.0 | -43 58 55 | 7.650 | 8.084 | 3.758 | 0.016 | 3.306 | 0.011 | 0.452 | 13.1801 |
| M | 8032 789 1 | 01 14 40.5 | -45 16 04 | 8.902 | 9.330 | 2.803 | 0.017 | 2.350 | 0.013 | 0.453 | 4.8257 |

| K, L, M | TYC | RA-2000 | Dec-2000 | V | B | bt | σBT | vt | σVT | χ | Plx |
|---|---|---|---|---|---|---|---|---|---|---|---|
| L | 511 234 1 | 20 42 23.59 | 00 26 42.1 | 9.088 | 9.502 | 2.379 | 0.022 | 1.926 | 0.018 | 0.453 | 3.6306 |
| M | 7540 618 1 | 01 18 22.6 | -44 11 49 | 8.613 | 9.053 | 3.986 | 0.016 | 3.531 | 0.012 | 0.455 | 9.4139 |
| M | 8446 398 1 | 22 38 04.5 | -46 42 35 | 7.680 | 8.109 | 3.053 | 0.017 | 2.597 | 0.011 | 0.456 | 9.4058 |
| M | 7959 931 1 | 20 00 20.7 | -44 40 50 | 9.134 | 9.572 | 3.008 | 0.021 | 2.546 | 0.016 | 0.462 | 4.6669 |
| M | 7895 1071 1 | 17 27 32.7 | -43 32 36 | 7.254 | 7.684 | 3.948 | 0.016 | 3.486 | 0.011 | 0.462 | 17.1636 |
| M | 8284 783 1 | 14 51 30.7 | -47 05 00 | 7.996 | 8.430 | 3.901 | 0.015 | 3.437 | 0.012 | 0.464 | 12.0498 |
| L | 4896 1560 1 | 09 54 18.60 | -00 15 06.9 | 9.962 | 10.389 | 3.440 | 0.037 | 2.974 | 0.034 | 0.466 | 3.9272 |
| M | 9412 2493 1 | 12 19 45.4 | -76 48 04 | 6.842 | 7.280 | 3.590 | 0.015 | 3.121 | 0.010 | 0.469 | 17.6322 |
| M | 9366 741 1 | 03 15 57.7 | -77 23 18 | 5.516 | 5.951 | 3.692 | 0.014 | 3.223 | 0.009 | 0.469 | 34.1190 |
| M | 8033 657 1 | 01 19 08.3 | -46 37 46 | 8.762 | 9.210 | 3.925 | 0.017 | 3.452 | 0.013 | 0.473 | 8.4718 |
| M | 8117 1539 1 | 06 49 54.6 | -46 36 52 | 5.141 | 5.592 | 3.649 | 0.014 | 3.175 | 0.009 | 0.474 | 39.5900 |
| M | 9318 88 1 | 19 59 22.3 | -74 07 49 | 7.514 | 7.949 | 2.400 | 0.015 | 1.922 | 0.010 | 0.478 | 7.4513 |
| M | 8886 240 1 | 05 12 03.1 | -65 10 33 | 7.037 | 7.471 | 3.608 | 0.015 | 3.130 | 0.010 | 0.478 | 16.1267 |
| M | 8229 1104 1 | 12 00 14.5 | -45 34 36 | 9.766 | 10.261 | 3.384 | 0.024 | 2.905 | 0.021 | 0.479 | 4.1366 |
| M | 7959 1763 1 | 20 08 01.3 | -44 11 34 | 8.476 | 8.924 | 4.191 | 0.017 | 3.712 | 0.013 | 0.479 | 10.8942 |
| M | 9364 933 1 | 04 16 14.7 | -75 20 39 | 8.465 | 8.893 | 1.782 | 0.017 | 1.299 | 0.013 | 0.483 | 3.6347 |
| M | 7768 2097 1 | 12 11 28.6 | -44 16 58 | 6.581 | 7.039 | 3.998 | 0.015 | 3.513 | 0.009 | 0.485 | 23.7845 |
| M | 8409 1949 1 | 20 19 17.9 | -47 34 49 | 6.122 | 6.590 | 4.179 | 0.014 | 3.694 | 0.010 | 0.485 | 31.7524 |
| M | 9461 1989 1 | 20 42 03.0 | -76 10 50 | 5.972 | 6.418 | 1.466 | 0.014 | 0.980 | 0.009 | 0.486 | 9.7726 |
| M | 7832 2494 1 | 14 59 45.0 | -43 48 41 | 6.582 | 7.053 | 4.311 | 0.015 | 3.818 | 0.009 | 0.493 | 27.4303 |
| M | 8104 1196 1 | 06 40 54.5 | -45 44 47 | 9.650 | 10.104 | 3.338 | 0.027 | 2.842 | 0.025 | 0.496 | 4.2078 |
| M | 8279 16 1 | 14 45 46.2 | -46 48 12 | 8.664 | 9.114 | 3.095 | 0.017 | 2.598 | 0.013 | 0.497 | 5.9419 |
| M | 8446 247 1 | 22 40 51.3 | -45 12 15 | 8.124 | 8.602 | 4.118 | 0.017 | 3.621 | 0.013 | 0.497 | 12.2133 |
| M | 7626 1877 1 | 06 37 58.1 | -44 17 03 | 8.464 | 8.928 | 4.120 | 0.018 | 3.621 | 0.013 | 0.499 | 10.4582 |
| L | 4914 343 1 | 10 53 11.45 | -01 02 42.7 | 9.910 | 10.402 | 4.397 | 0.031 | 3.897 | 0.028 | 0.500 | 6.1074 |
| M | 8446 948 1 | 22 31 47.8 | -46 34 47 | 7.233 | 7.693 | 2.076 | 0.015 | 1.576 | 0.010 | 0.500 | 7.2059 |
| M | 8391 1625 1 | 19 58 58.5 | -46 05 17 | 7.474 | 7.951 | 4.466 | 0.015 | 3.960 | 0.011 | 0.506 | 19.3258 |
| M | 8889 314 1 | 04 58 58.2 | -66 12 15 | 9.805 | 10.334 | 3.884 | 0.037 | 3.375 | 0.032 | 0.509 | 4.8798 |
| M | 9139 344 1 | 01 13 03.7 | -71 38 03 | 9.775 | 10.212 | 3.909 | 0.029 | 3.398 | 0.026 | 0.511 | 5.1069 |
| M | 8067 281 1 | 03 57 51.7 | -46 22 59 | 8.011 | 8.489 | 4.350 | 0.017 | 3.839 | 0.012 | 0.511 | 14.2441 |
| M | 8279 693 1 | 14 46 37.7 | -45 07 23 | 9.958 | 10.408 | 3.383 | 0.033 | 2.871 | 0.020 | 0.512 | 3.7874 |
| M | 7959 547 1 | 20 06 03.0 | -44 58 49 | 8.063 | 8.531 | 3.373 | 0.016 | 2.861 | 0.012 | 0.512 | 8.9047 |
| M | 7703 532 1 | 09 29 22.5 | -43 11 25 | 6.601 | 7.077 | 2.117 | 0.015 | 1.603 | 0.010 | 0.514 | 9.7793 |
| M | 8229 295 1 | 12 00 04.9 | -46 02 49 | 9.820 | 10.298 | 4.612 | 0.024 | 4.096 | 0.021 | 0.516 | 7.0000 |
| M | 8279 692 1 | 14 42 49.1 | -45 35 51 | 8.130 | 8.622 | -1.295 | 0.016 | -1.811 | 0.012 | 0.516 | 1.0024 |
| M | 7583 1599 1 | 04 01 08.5 | -44 26 44 | 8.464 | 8.960 | 4.585 | 0.017 | 4.067 | 0.013 | 0.518 | 12.7981 |
| M | 9141 7665 1 | 00 38 17.8 | -73 57 46 | 7.437 | 7.918 | 4.074 | 0.016 | 3.556 | 0.011 | 0.518 | 16.3705 |
| M | 8229 2132 1 | 12 10 53.6 | -45 33 46 | 10.220 | 10.656 | 1.512 | 0.031 | 0.994 | 0.026 | 0.518 | 1.4373 |
| M | 8166 533 1 | 09 14 10.9 | -45 03 59 | 8.505 | 8.994 | 3.973 | 0.017 | 3.453 | 0.013 | 0.520 | 9.4695 |
| M | 7704 2481 1 | 09 30 19.4 | -44 51 35 | 7.570 | 8.058 | 2.815 | 0.016 | 2.292 | 0.011 | 0.523 | 8.5873 |
| L | 543 1152 1 | 21 41 56.46 | 00 20 45.7 | 7.653 | 8.141 | 3.093 | 0.017 | 2.568 | 0.012 | 0.525 | 9.3457 |
| L | 543 227 1 | 21 41 27.38 | 00 40 15.5 | 10.007 | 10.459 | 3.074 | 0.045 | 2.548 | 0.039 | 0.526 | 3.1870 |
| L | 66 772 1 | 03 54 15.29 | 00 17 22.1 | 8.737 | 9.239 | 3.498 | 0.020 | 2.971 | 0.016 | 0.527 | 6.8015 |

Conversion of Tycho-2 to Johnson-Cousins magnitudes in the Gaia era

| K, L, M | TYC | RA-2000 | Dec-2000 | V | B | bt | σBT | vt | σVT | χ | Plx |
|---|---|---|---|---|---|---|---|---|---|---|---|
| M | 7895 9 1 | 17 29 15.3 | -43 30 40 | 9.739 | 10.285 | 2.445 | 0.044 | 1.917 | 0.044 | 0.528 | 2.6331 |
| M | 8011 470 1 | 22 55 25.2 | -42 53 04 | 8.467 | 8.963 | 3.666 | 0.017 | 3.134 | 0.013 | 0.532 | 8.2719 |
| L | 30 52 1 | 01 55 03.17 | 00 40 08.8 | 9.789 | 10.307 | 4.401 | 0.034 | 3.866 | 0.030 | 0.535 | 6.3728 |
| M | 7814 30 1 | 14 31 41.4 | -43 05 18 | 7.857 | 8.349 | 3.399 | 0.016 | 2.860 | 0.011 | 0.539 | 9.7829 |
| M | 8877 184 1 | 05 03 18.2 | -61 24 55 | 8.757 | 9.267 | 3.591 | 0.020 | 3.052 | 0.015 | 0.539 | 7.0942 |
| M | 8010 303 1 | 22 50 41.4 | -43 36 21 | 9.587 | 10.151 | 4.792 | 0.031 | 4.252 | 0.027 | 0.540 | 8.2065 |
| M | 8446 467 1 | 22 43 37.8 | -45 25 03 | 10.345 | 10.861 | 3.249 | 0.048 | 2.709 | 0.041 | 0.540 | 2.9498 |
| L | 4896 1258 1 | 09 57 05.54 | -00 30 03.0 | 10.002 | 10.431 | 3.563 | 0.048 | 3.022 | 0.042 | 0.541 | 3.9424 |
| M | 7540 885 1 | 01 15 10.4 | -44 37 17 | 8.369 | 8.861 | 2.658 | 0.016 | 2.116 | 0.011 | 0.542 | 5.4837 |
| M | 8279 3080 1 | 14 39 10.8 | -46 35 04 | 6.063 | 6.570 | 3.991 | 0.014 | 3.449 | 0.009 | 0.542 | 29.3708 |
| M | 7576 389 1 | 03 56 17.7 | -43 37 05 | 8.131 | 8.636 | 4.032 | 0.017 | 3.485 | 0.012 | 0.547 | 11.4652 |
| M | 8104 1052 1 | 06 42 31.3 | -45 21 21 | 9.530 | 9.993 | 2.199 | 0.027 | 1.649 | 0.022 | 0.550 | 2.5996 |
| L | 30 193 1 | 01 54 50.38 | 00 47 00.8 | 9.569 | 10.023 | 3.833 | 0.028 | 3.277 | 0.024 | 0.556 | 5.5520 |
| M | 7574 924 1 | 01 31 32.7 | -43 50 48 | 8.187 | 8.705 | 1.428 | 0.016 | 0.869 | 0.011 | 0.559 | 3.3538 |
| M | 7639 1690 1 | 06 45 14.4 | -44 27 17 | 9.727 | 10.195 | 4.218 | 0.028 | 3.658 | 0.023 | 0.560 | 6.0385 |
| L | 48 903 1 | 02 56 13.83 | 00 31 20.0 | 8.743 | 9.237 | 3.352 | 0.028 | 2.791 | 0.022 | 0.561 | 6.2935 |
| M | 8220 491 1 | 11 58 15.6 | -47 58 38 | 6.766 | 7.288 | 4.507 | 0.015 | 3.944 | 0.010 | 0.563 | 26.5912 |
| M | 9412 1917 1 | 12 13 54.2 | -76 01 19 | 8.686 | 9.234 | 2.615 | 0.018 | 2.052 | 0.014 | 0.563 | 4.5487 |
| M | 9319 933 1 | 20 09 04.2 | -74 13 10 | 7.593 | 8.108 | 3.934 | 0.016 | 3.370 | 0.011 | 0.564 | 13.9109 |
| M | 8003 1371 1 | 22 36 42.6 | -43 28 16 | 6.756 | 7.281 | 3.762 | 0.015 | 3.195 | 0.010 | 0.567 | 18.7818 |
| M | 8167 639 1 | 09 23 30.8 | -45 32 43 | 10.003 | 10.552 | 2.986 | 0.037 | 2.419 | 0.031 | 0.567 | 2.9619 |
| M | 9411 33 1 | 11 42 14.9 | -75 13 38 | 6.471 | 6.995 | 4.435 | 0.015 | 3.865 | 0.010 | 0.570 | 29.3542 |
| L | 4966 992 1 | 13 36 14.64 | -00 55 51.5 | 7.062 | 7.590 | 4.171 | 0.016 | 3.589 | 0.011 | 0.582 | 19.6782 |
| M | 8891 2444 1 | 05 31 37.4 | -66 40 41 | 8.876 | 9.420 | 4.265 | 0.021 | 3.681 | 0.015 | 0.584 | 8.8618 |
| M | 7767 255 1 | 12 00 09.0 | -44 51 38 | 7.850 | 8.392 | 3.053 | 0.015 | 2.469 | 0.011 | 0.584 | 8.1568 |
| M | 9138 1892 1 | 00 50 12.6 | -71 29 35 | 9.567 | 10.119 | 3.229 | 0.029 | 2.644 | 0.023 | 0.585 | 3.9341 |
| M | 9460 411 1 | 20 02 50.1 | -75 20 58 | 8.220 | 8.752 | 4.150 | 0.016 | 3.563 | 0.011 | 0.587 | 11.2937 |
| M | 8279 1790 1 | 14 40 11.1 | -45 37 48 | 9.943 | 10.454 | 4.920 | 0.036 | 4.332 | 0.029 | 0.588 | 7.3068 |
| M | 7832 1250 1 | 14 52 15.5 | -44 55 38 | 7.206 | 7.746 | 2.779 | 0.014 | 2.190 | 0.010 | 0.589 | 9.6796 |
| M | 9459 663 1 | 19 43 00.0 | -75 23 51 | 8.006 | 8.546 | 3.729 | 0.015 | 3.138 | 0.011 | 0.591 | 10.3026 |
| M | 9138 70 1 | 00 44 00.8 | -72 21 49 | 9.776 | 10.236 | 4.626 | 0.030 | 4.033 | 0.024 | 0.593 | 7.0025 |
| M | 8117 1258 1 | 06 52 27.9 | -45 02 56 | 8.847 | 9.407 | 4.988 | 0.019 | 4.395 | 0.015 | 0.593 | 12.5942 |
| M | 7576 580 1 | 03 58 10.5 | -44 00 17 | 10.312 | 10.782 | 3.769 | 0.045 | 3.172 | 0.036 | 0.597 | 3.7082 |
| L | 4914 1028 1 | 10 55 25.37 | -00 48 46.8 | 8.890 | 9.442 | 4.557 | 0.020 | 3.960 | 0.015 | 0.597 | 10.1099 |
| M | 8280 1620 1 | 14 52 23.7 | -46 37 50 | 8.996 | 9.530 | 2.294 | 0.021 | 1.695 | 0.017 | 0.599 | 3.3905 |
| M | 8229 2125 1 | 12 10 57.9 | -46 19 19 | 7.528 | 8.088 | 5.416 | 0.016 | 4.816 | 0.010 | 0.600 | 27.9076 |
| M | 8216 2579 1 | 11 59 35.3 | -46 37 59 | 6.655 | 7.210 | 2.444 | 0.015 | 1.842 | 0.010 | 0.602 | 10.6073 |
| L | 5017 78 1 | 15 39 01.05 | -00 18 41.4 | 7.500 | 8.043 | 4.630 | 0.017 | 4.028 | 0.011 | 0.602 | 19.6613 |
| M | 7891 262 1 | 17 22 30.1 | -42 57 29 | 9.822 | 10.459 | 3.821 | 0.039 | 3.212 | 0.032 | 0.609 | 4.4578 |
| M | 8010 806 1 | 22 47 13.9 | -43 36 22 | 8.807 | 9.357 | 4.590 | 0.019 | 3.979 | 0.015 | 0.611 | 10.4880 |
| M | 7547 638 1 | 01 27 30.6 | -43 56 03 | 9.855 | 10.434 | 4.586 | 0.027 | 3.973 | 0.023 | 0.613 | 6.4375 |
| M | 8033 1232 1 | 01 15 11.1 | -45 31 54 | 4.966 | 5.536 | 4.750 | 0.014 | 4.135 | 0.009 | 0.615 | 66.43 |
| M | 8284 505 1 | 14 56 34.4 | -46 57 02 | 7.089 | 7.669 | 0.435 | 0.015 | -0.183 | 0.010 | 0.618 | 3.4068 |

| K, L, M | TYC | RA-2000 | Dec-2000 | V | B | bt | σBT | vt | σVT | χ | Plx |
|---|---|---|---|---|---|---|---|---|---|---|---|
| M | 9142 2820 1 | 01 05 18.2 | -72 33 14 | 8.063 | 8.623 | 4.843 | 0.017 | 4.223 | 0.012 | 0.620 | 16.6654 |
| M | 279 2118 1 | 14 40 12.7 | -45 32 45 | 8.118 | 8.686 | 4.820 | 0.017 | 4.188 | 0.012 | 0.632 | 15.9620 |
| L | 5048 708 1 | 16 35 57.49 | -00 24 41.4 | 8.208 | 8.767 | 5.026 | 0.017 | 4.394 | 0.012 | 0.632 | 16.7142 |
| M | 7583 529 1 | 04 07 06.2 | -44 56 44 | 8.345 | 8.915 | 5.082 | 0.017 | 4.449 | 0.013 | 0.633 | 16.1603 |
| L | 5082 1171 1 | 17 45 42.44 | -00 21 35.4 | 10.358 | 10.966 | 4.497 | 0.056 | 3.864 | 0.046 | 0.633 | 4.7402 |
| M | 7547 913 1 | 01 21 52.7 | -44 39 37 | 9.756 | 10.352 | 3.192 | 0.027 | 2.558 | 0.021 | 0.634 | 3.5460 |
| M | 8117 420 1 | 06 45 48.5 | -45 59 52 | 8.996 | 9.564 | 4.964 | 0.022 | 4.327 | 0.017 | 0.637 | 11.3548 |
| M | 8391 1305 1 | 19 56 58.1 | -45 49 00 | 8.812 | 9.375 | 1.998 | 0.019 | 1.353 | 0.014 | 0.645 | 3.1453 |
| M | 7686 3000 1 | 09 13 44.6 | -42 18 37 | 8.017 | 8.590 | 5.036 | 0.016 | 4.390 | 0.011 | 0.646 | 18.3705 |
| M | 7832 2455 1 | 14 59 27.2 | -43 09 35 | 6.104 | 6.700 | -0.466 | 0.015 | -1.115 | 0.010 | 0.649 | 3.5022 |
| M | 8229 1054 1 | 12 00 07.2 | -45 32 30 | 9.940 | 10.548 | 4.690 | 0.031 | 4.040 | 0.024 | 0.650 | 6.5343 |
| M | 7583 1011 1 | 04 11 44.8 | -44 49 02 | 9.680 | 10.251 | 4.808 | 0.027 | 4.157 | 0.023 | 0.651 | 7.7344 |
| M | 7956 817 1 | 20 18 39.2 | -42 37 13 | 7.450 | 8.040 | 3.335 | 0.017 | 2.681 | 0.011 | 0.654 | 10.7849 |
| M | 8168 947 1 | 09 36 38.4 | -46 20 19 | 8.393 | 8.979 | 5.165 | 0.017 | 4.507 | 0.012 | 0.658 | 16.1360 |
| M | 7639 1742 1 | 06 40 51.4 | -44 24 56 | 8.674 | 9.300 | 5.190 | 0.019 | 4.528 | 0.014 | 0.662 | 14.3189 |
| L | 84 1220 1 | 04 52 29.68 | 00 02 06.3 | 9.652 | 10.250 | 5.385 | 0.035 | 4.722 | 0.031 | 0.663 | 10.0929 |
| M | 8447 665 1 | 22 54 16.2 | -45 53 19 | 8.296 | 8.880 | 5.474 | 0.019 | 4.804 | 0.014 | 0.670 | 19.5843 |
| L | 117 759 1 | 05 57 27.29 | 00 13 20.4 | 9.260 | 9.854 | 4.054 | 0.030 | 3.376 | 0.022 | 0.678 | 6.5992 |
| L | 5016 518 1 | 15 37 18.14 | -00 09 49.6 | 8.433 | 9.052 | 4.309 | 0.020 | 3.628 | 0.015 | 0.681 | 10.5664 |
| M | 9461 341 1 | 20 43 13.9 | -76 32 51 | 7.660 | 8.265 | 4.292 | 0.016 | 3.607 | 0.010 | 0.685 | 14.9589 |
| M | 8069 1066 1 | 04 18 00.1 | -45 39 04 | 7.535 | 8.175 | 4.506 | 0.016 | 3.807 | 0.010 | 0.699 | 17.3292 |
| M | 8117 1888 1 | 06 52 23.7 | -46 02 53 | 8.330 | 8.974 | 4.104 | 0.017 | 3.402 | 0.013 | 0.702 | 10.0100 |
| M | 7892 7606 1 | 17 37 53.9 | -42 34 03 | 7.165 | 7.813 | 5.271 | 0.017 | 4.558 | 0.011 | 0.713 | 29.2433 |
| M | 8229 2279 1 | 12 00 03.4 | -46 47 03 | 8.378 | 9.030 | 4.710 | 0.016 | 3.990 | 0.012 | 0.720 | 12.9423 |
| L | 586 979 1 | 23 42 41.82 | 00 45 13.1 | 9.695 | 10.314 | 3.538 | 0.042 | 2.816 | 0.032 | 0.722 | 4.1462 |
| M | 8171 686 1 | 09 30 33.1 | -47 36 16 | 8.649 | 9.297 | 5.825 | 0.018 | 5.102 | 0.013 | 0.723 | 18.9489 |
| M | 7583 1551 1 | 04 02 58.4 | -44 23 32 | 8.124 | 8.770 | 4.680 | 0.017 | 3.954 | 0.013 | 0.726 | 14.2816 |
| M | 8234 2891 1 | 12 15 36.7 | -48 24 58 | 8.940 | 9.610 | 5.794 | 0.019 | 5.067 | 0.014 | 0.727 | 16.3139 |
| L | 4932 454 1 | 11 53 59.74 | -00 31 46.5 | 10.111 | 10.675 | 3.382 | 0.040 | 2.635 | 0.028 | 0.747 | 3.1399 |
| M | 8279 897 1 | 14 45 45.5 | -45 52 52 | 8.479 | 9.165 | 4.789 | 0.017 | 4.042 | 0.012 | 0.747 | 12.3254 |
| M | 7689 759 1 | 09 06 38.8 | -43 29 31 | 7.306 | 7.978 | 5.208 | 0.015 | 4.457 | 0.010 | 0.751 | 26.1438 |
| M | 7768 1998 1 | 12 18 36.1 | -44 08 42 | 8.410 | 9.096 | 5.914 | 0.017 | 5.147 | 0.012 | 0.767 | 21.6250 |
| M | 8346 41 1 | 17 27 46.4 | -46 57 04 | 8.153 | 8.851 | 4.283 | 0.017 | 3.512 | 0.012 | 0.771 | 11.3191 |
| L | 511 1837 1 | 20 41 35.82 | 00 16 35.8 | 9.853 | 10.541 | 4.354 | 0.040 | 3.577 | 0.030 | 0.777 | 5.4554 |
| M | 7635 1587 1 | 06 44 01.4 | -42 42 52 | 8.534 | 9.232 | 5.806 | 0.019 | 5.021 | 0.014 | 0.785 | 19.3417 |
| L | 4985 376 1 | 14 39 28.28 | -00 14 56.3 | 9.088 | 9.789 | 5.194 | 0.029 | 4.399 | 0.021 | 0.795 | 11.2055 |
| M | 8278 2535 1 | 14 29 03.3 | -46 44 28 | 8.932 | 9.616 | 6.247 | 0.022 | 5.448 | 0.014 | 0.799 | 19.6955 |
| M | 7896 3839 1 | 17 38 59.5 | -43 08 44 | 7.245 | 7.967 | 4.401 | 0.019 | 3.599 | 0.013 | 0.802 | 17.8763 |
| L | 416 537 1 | 17 45 29.89 | 00 04 12.0 | 9.017 | 9.819 | 5.055 | 0.028 | 4.240 | 0.019 | 0.815 | 10.6277 |
| M | 7626 488 1 | 06 37 25.9 | -43 39 40 | 9.926 | 10.717 | 6.799 | 0.046 | 5.973 | 0.032 | 0.826 | 15.2837 |
| M | 7547 945 1 | 01 23 16.7 | -44 40 18 | 9.852 | 10.597 | 3.626 | 0.034 | 2.798 | 0.024 | 0.828 | 3.6787 |
| M | 8069 1348 1 | 04 15 45.2 | -45 58 10 | 8.922 | 9.695 | 6.431 | 0.024 | 5.592 | 0.016 | 0.839 | 20.5991 |
| M | 8007 697 1 | 22 50 10.5 | -41 29 24 | 7.770 | 8.514 | 6.423 | 0.017 | 5.581 | 0.012 | 0.842 | 35.0508 |

Conversion of Tycho-2 to Johnson-Cousins magnitudes in the Gaia era

| K, L, M | TYC | RA-2000 | Dec-2000 | V | B | bt | σBT | vt | σVT | χ | Plx |
|---|---|---|---|---|---|---|---|---|---|---|---|
| M | 7547 886 1 | 01 26 16.7 | -44 41 30 | 9.638 | 10.462 | 3.179 | 0.032 | 2.334 | 0.022 | 0.845 | 3.2809 |
| M | 8229 1945 1 | 12 07 56.6 | -45 55 32 | 9.372 | 10.173 | 1.772 | 0.024 | 0.887 | 0.016 | 0.885 | 1.9358 |
| L | 383 1383 1 | 16 37 45.85 | 00 02 24.6 | 8.050 | 8.808 | 6.006 | 0.018 | 5.107 | 0.013 | 0.899 | 24.9595 |
| M | 8007 788 1 | 22 45 57.2 | -42 25 21 | 9.790 | 10.602 | 6.629 | 0.047 | 5.708 | 0.032 | 0.921 | 14.6950 |
| M | 7959 1661 1 | 20 01 41.9 | -44 28 10 | 9.396 | 10.277 | 3.008 | 0.032 | 2.062 | 0.020 | 0.946 | 3.2096 |
| M | 7583 1230 1 | 04 03 05.7 | -44 11 13 | 9.944 | 10.816 | 4.805 | 0.049 | 3.839 | 0.031 | 0.966 | 5.8916 |
| M | 8010 1053 1 | 22 51 09.1 | -44 44 12 | 8.430 | 9.334 | 3.216 | 0.021 | 2.217 | 0.014 | 0.999 | 5.4268 |
| M | 7960 524 1 | 20 11 15.8 | -43 39 44 | 6.542 | 7.427 | 1.192 | 0.015 | 0.182 | 0.010 | 1.010 | 5.1185 |
| M | 7576 189 1 | 03 50 35.5 | -42 33 56 | 8.619 | 9.541 | 7.611 | 0.021 | 6.581 | 0.013 | 1.030 | 37.4500 |
| M | 8167 62 1 | 09 31 03.5 | -45 27 45 | 8.797 | 9.707 | 1.524 | 0.024 | 0.490 | 0.015 | 1.034 | 2.0846 |
| M | 9142 2815 1 | 01 08 06.8 | -73 00 48 | 9.096 | 10.008 | 2.984 | 0.029 | 1.948 | 0.018 | 1.036 | 3.5936 |
| M | 8167 1048 1 | 09 29 05.3 | -45 27 25 | 9.078 | 10.005 | 1.324 | 0.027 | 0.280 | 0.016 | 1.044 | 1.6655 |
| M | 8889 524 1 | 04 58 16.6 | -66 15 36 | 9.274 | 10.197 | 2.021 | 0.034 | 0.975 | 0.021 | 1.046 | 2.1195 |
| M | 7832 701 1 | 14 52 17.7 | -43 49 05 | 8.707 | 9.643 | 7.233 | 0.025 | 6.179 | 0.015 | 1.054 | 30.1101 |
| M | 8230 1672 1 | 12 12 39.1 | -45 31 44 | 8.896 | 9.816 | 2.304 | 0.021 | 1.249 | 0.012 | 1.055 | 2.8373 |
| M | 7831 488 1 | 14 44 59.9 | -44 39 54 | 7.522 | 8.461 | 2.170 | 0.019 | 1.113 | 0.012 | 1.057 | 5.0110 |
| M | 8108 1387 1 | 06 44 13.5 | -47 04 33 | 8.168 | 9.130 | 1.921 | 0.021 | 0.861 | 0.013 | 1.060 | 3.2586 |
| M | 8117 1681 1 | 06 52 40.1 | -46 48 34 | 8.044 | 9.000 | 1.197 | 0.019 | 0.137 | 0.013 | 1.060 | 2.5050 |
| M | 8171 95 1 | 09 30 59.3 | -47 23 03 | 8.502 | 9.412 | 7.193 | 0.021 | 6.130 | 0.013 | 1.063 | 31.7697 |
| M | 7959 1599 1 | 20 02 34.4 | -44 13 19 | 8.897 | 9.869 | 2.426 | 0.025 | 1.359 | 0.015 | 1.067 | 2.9263 |
| M | 8067 1853 1 | 03 50 02.6 | -45 23 07 | 6.934 | 7.881 | 1.707 | 0.015 | 0.630 | 0.010 | 1.077 | 5.2359 |
| M | 8003 1022 1 | 22 33 54.7 | -44 41 25 | 6.785 | 7.731 | 1.912 | 0.016 | 0.817 | 0.010 | 1.095 | 6.1124 |
| M | 8216 796 1 | 11 54 23.6 | -46 29 30 | 7.975 | 8.957 | 2.767 | 0.017 | 1.670 | 0.011 | 1.097 | 5.2455 |
| L | 48 865 1 | 02 58 28.69 | 00 26 14.1 | 9.041 | 10.034 | 2.304 | 0.049 | 1.197 | 0.029 | 1.107 | 2.4959 |
| M | 8280 2302 1 | 14 55 19.0 | -46 37 53 | 7.314 | 8.287 | 1.911 | 0.016 | 0.803 | 0.010 | 1.108 | 4.7955 |
| M | 9160 1808 1 | 04 34 15.4 | -73 12 32 | 6.807 | 7.770 | 1.273 | 0.016 | 0.164 | 0.010 | 1.109 | 4.4711 |
| M | 8278 3082 1 | 14 35 50.0 | -46 48 57 | 6.782 | 7.744 | 1.212 | 0.015 | 0.103 | 0.010 | 1.109 | 4.4163 |
| M | 8342 2016 1 | 17 34 40.1 | -45 07 54 | 8.748 | 9.694 | 2.466 | 0.032 | 1.351 | 0.019 | 1.115 | 3.0992 |
| M | 7547 1103 1 | 01 29 34.0 | -44 48 33 | 9.347 | 10.297 | 5.140 | 0.030 | 4.022 | 0.017 | 1.118 | 8.3043 |
| M | 8166 1466 1 | 09 18 33.4 | -45 32 07 | 9.302 | 10.286 | 2.021 | 0.033 | 0.901 | 0.018 | 1.120 | 1.9859 |
| M | 8010 468 1 | 22 47 08.8 | -43 53 46 | 8.712 | 9.715 | 1.751 | 0.024 | 0.629 | 0.015 | 1.122 | 2.2787 |
| M | 7960 2048 1 | 20 18 56.6 | -44 31 28 | 7.551 | 8.513 | 1.996 | 0.017 | 0.871 | 0.011 | 1.125 | 4.4250 |
| M | 8010 1573 1 | 22 42 43.1 | -44 14 53 | 6.065 | 7.031 | 2.858 | 0.015 | 1.733 | 0.010 | 1.125 | 12.9285 |
| M | 8043 1179 1 | 01 31 15.1 | -49 04 22 | 3.944 | 4.935 | 2.017 | 0.014 | 0.885 | 0.009 | 1.132 | 23.3497 |
| M | 7626 763 1 | 06 31 04.5 | -43 43 01 | 6.689 | 7.689 | 1.954 | 0.016 | 0.807 | 0.010 | 1.147 | 6.3477 |
| M | 8167 1799 1 | 09 25 13.1 | -45 29 01 | 8.845 | 9.829 | 2.883 | 0.025 | 1.736 | 0.014 | 1.147 | 3.5996 |
| M | 7704 2608 1 | 09 38 01.5 | -43 11 27 | 5.498 | 6.500 | 0.837 | 0.015 | -0.314 | 0.009 | 1.151 | 6.5600 |
| M | 8392 1069 1 | 20 08 44.8 | -45 25 43 | 9.040 | 10.050 | 1.988 | 0.030 | 0.832 | 0.017 | 1.156 | 2.1769 |
| M | 8003 474 1 | 22 32 11.5 | -43 15 47 | 6.905 | 7.890 | 2.016 | 0.016 | 0.860 | 0.010 | 1.156 | 5.8814 |
| M | 8010 937 1 | 22 48 30.2 | -43 26 47 | 7.064 | 8.053 | 1.872 | 0.016 | 0.716 | 0.010 | 1.156 | 5.1292 |
| M | 8229 1505 1 | 12 02 49.8 | -46 23 19 | 9.492 | 10.536 | 1.616 | 0.033 | 0.459 | 0.018 | 1.157 | 1.4927 |
| M | 7957 1099 1 | 20 20 22.6 | -42 54 04 | 7.764 | 8.788 | 2.132 | 0.019 | 0.973 | 0.012 | 1.159 | 4.1533 |
| M | 9139 2213 1 | 01 04 21.0 | -72 15 09 | 8.421 | 9.432 | 2.081 | 0.020 | 0.919 | 0.012 | 1.162 | 2.9858 |

| K, L, M | TYC | RA-2000 | Dec-2000 | V | B | bt | σBT | vt | σVT | χ | Plx |
|---|---|---|---|---|---|---|---|---|---|---|---|
| M | 7583 851 1 | 04 03 04.8 | -44 35 31 | 8.755 | 9.773 | 3.842 | 0.029 | 2.679 | 0.016 | 1.163 | 5.8526 |
| M | 8002 1698 1 | 22 15 35.0 | -44 27 05 | 6.102 | 7.117 | 0.482 | 0.015 | -0.681 | 0.010 | 1.163 | 4.2055 |
| M | 8040 737 1 | 01 37 45.8 | -47 10 41 | 7.579 | 8.601 | 1.725 | 0.016 | 0.560 | 0.011 | 1.165 | 3.7626 |
| M | 8033 210 1 | 01 17 43.8 | -46 10 21 | 8.010 | 9.016 | 3.449 | 0.017 | 2.280 | 0.011 | 1.169 | 6.7722 |
| M | 8278 632 1 | 14 31 28.2 | -45 42 52 | 9.028 | 10.067 | 2.040 | 0.030 | 0.870 | 0.017 | 1.170 | 2.2187 |
| L | 12 104 1 | 00 55 01.4 | 00 47 22.8 | 8.046 | 9.031 | 3.904 | 0.021 | 2.730 | 0.013 | 1.174 | 8.2083 |
| M | 8229 2807 1 | 12 10 39.9 | -46 31 45 | 8.347 | 9.374 | 2.105 | 0.019 | 0.930 | 0.012 | 1.175 | 3.1218 |
| M | 8170 165 1 | 09 19 00.2 | -47 58 04 | 7.026 | 8.045 | 4.141 | 0.016 | 2.964 | 0.010 | 1.177 | 14.6343 |
| M | 8010 425 1 | 22 46 36.0 | -43 53 09 | 9.230 | 10.240 | 2.006 | 0.034 | 0.825 | 0.019 | 1.181 | 1.9941 |
| M | 8104 1227 1 | 06 39 42.4 | -45 33 38 | 8.584 | 9.601 | 1.996 | 0.024 | 0.811 | 0.014 | 1.185 | 2.6738 |
| L | 326 981 1 | 14 41 26.70 | 00 06 12.5 | 8.123 | 9.152 | 2.075 | 0.020 | 0.889 | 0.013 | 1.186 | 3.3946 |
| M | 7960 1218 1 | 20 10 05.9 | -43 53 39 | 6.928 | 7.954 | 1.994 | 0.016 | 0.808 | 0.010 | 1.186 | 5.6895 |
| M | 7544 1147 1 | 01 24 40.8 | -41 29 33 | 5.422 | 6.468 | 2.255 | 0.014 | 1.066 | 0.009 | 1.189 | 12.7929 |
| M | 8216 536 1 | 11 53 36.0 | -46 44 30 | 7.222 | 8.260 | 2.272 | 0.015 | 1.079 | 0.010 | 1.193 | 5.6174 |
| M | 8068 1502 1 | 04 01 17.6 | -46 08 14 | 7.237 | 8.345 | 1.818 | 0.017 | 0.622 | 0.010 | 1.196 | 4.5235 |
| L | 4736 933 1 | 04 51 40.76 | -00 05 13.1 | 8.930 | 9.979 | 1.508 | 0.033 | 0.311 | 0.019 | 1.197 | 1.7734 |
| M | 7547 152 1 | 01 29 26.7 | -43 09 24 | 8.303 | 9.343 | 1.954 | 0.018 | 0.755 | 0.011 | 1.199 | 2.9534 |
| L | 4914 464 1 | 10 54 50.45 | -00 55 05.5 | 8.754 | 9.768 | 1.968 | 0.026 | 0.764 | 0.014 | 1.204 | 2.3845 |
| M | 7576 345 1 | 03 58 52.5 | -44 29 53 | 9.394 | 10.434 | 5.113 | 0.033 | 3.908 | 0.019 | 1.205 | 7.7007 |
| M | 9315 1438 1 | 20 18 30.0 | -72 48 29 | 6.930 | 7.961 | 1.894 | 0.015 | 0.688 | 0.010 | 1.206 | 5.3521 |
| M | 9319 1078 1 | 20 00 56.0 | -74 00 45 | 6.582 | 7.627 | 2.057 | 0.015 | 0.848 | 0.010 | 1.209 | 6.7764 |
| L | 4966 981 1 | 13 36 43.99 | -01 15 38.1 | 8.345 | 9.384 | 2.009 | 0.025 | 0.798 | 0.014 | 1.211 | 2.9215 |
| M | 7832 2261 1 | 14 54 00.8 | -43 35 52 | 8.218 | 9.278 | 0.939 | 0.020 | -0.281 | 0.012 | 1.220 | 1.8989 |
| M | 8010 1117 1 | 22 51 48.5 | -44 47 04 | 8.341 | 9.418 | 1.942 | 0.023 | 0.715 | 0.014 | 1.227 | 2.8351 |
| M | 8010 893 1 | 22 48 21.0 | -44 30 04 | 8.866 | 9.959 | 2.027 | 0.030 | 0.798 | 0.017 | 1.229 | 2.2929 |
| M | 7831 2339 1 | 14 47 32.1 | -43 33 25 | 6.304 | 7.385 | 0.775 | 0.016 | -0.472 | 0.010 | 1.247 | 4.1868 |
| M | 8405 624 1 | 20 17 40.3 | -45 13 44 | 8.565 | 9.661 | 2.281 | 0.027 | 1.034 | 0.015 | 1.247 | 2.9903 |
| M | 7959 865 1 | 20 03 58.1 | -44 38 19 | 8.338 | 9.421 | 2.153 | 0.021 | 0.905 | 0.013 | 1.248 | 3.1035 |
| M | 7959 511 1 | 20 02 59.3 | -44 50 53 | 9.402 | 10.496 | 2.594 | 0.040 | 1.342 | 0.021 | 1.252 | 2.3012 |
| M | 8103 1527 1 | 06 32 19.6 | -45 18 36 | 7.171 | 8.249 | 1.954 | 0.017 | 0.700 | 0.010 | 1.254 | 4.8085 |
| M | 9319 757 1 | 20 08 30.3 | -74 45 27 | 7.241 | 8.306 | 2.165 | 0.016 | 0.909 | 0.010 | 1.256 | 5.1057 |
| M | 7540 259 1 | 01 18 54.6 | -44 56 13 | 8.928 | 10.010 | 1.068 | 0.024 | -0.190 | 0.013 | 1.258 | 1.4414 |
| L | 4865 1177 1 | 08 53 58.81 | -00 36 43.2 | 8.409 | 9.527 | 2.482 | 0.026 | 1.223 | 0.015 | 1.259 | 3.3952 |
| M | 7942 3241 1 | 19 55 15.7 | -41 52 06 | 4.119 | 5.203 | 1.619 | 0.014 | 0.355 | 0.009 | 1.264 | 16.7487 |
| L | 4914 455 1 | 10 54 37.34 | -00 55 28.6 | 9.246 | 10.302 | 3.549 | 0.037 | 2.280 | 0.018 | 1.269 | 3.8195 |
| M | 7581 1600 1 | 04 14 00.1 | -42 17 40 | 3.853 | 4.950 | 2.477 | 0.014 | 1.196 | 0.009 | 1.281 | 27.9235 |
| M | 7548 607 1 | 01 34 00.1 | -44 23 48 | 8.545 | 9.622 | 2.058 | 0.023 | 0.770 | 0.013 | 1.288 | 2.6376 |
| M | 7576 1635 1 | 03 53 48.0 | -43 09 44 | 7.922 | 9.020 | 2.450 | 0.018 | 1.162 | 0.011 | 1.288 | 4.2214 |
| M | 8280 1641 1 | 14 52 43.0 | -45 35 26 | 9.142 | 10.230 | 2.247 | 0.036 | 0.959 | 0.018 | 1.288 | 2.1677 |
| L | 4896 633 1 | 09 55 35.14 | -01 07 34.6 | 7.997 | 9.105 | 2.450 | 0.022 | 1.161 | 0.013 | 1.289 | 4.0872 |
| M | 7959 175 1 | 20 08 29.2 | -44 39 07 | 9.518 | 10.644 | 2.269 | 0.047 | 0.979 | 0.024 | 1.290 | 1.8442 |
| M | 9368 84 1 | 04 38 21.7 | -77 39 22 | 6.054 | 7.147 | 1.371 | 0.015 | 0.078 | 0.010 | 1.293 | 6.0733 |

Conversion of Tycho-2 to Johnson-Cousins magnitudes in the Gaia era

| K, L, M | TYC | RA-2000 | Dec-2000 | V | B | bt | σBT | vt | σVT | χ | Plx |
|---|---|---|---|---|---|---|---|---|---|---|---|
| M | 9236 2693 1 | 12 32 10.0 | -73 00 04 | 5.881 | 6.984 | 2.010 | 0.015 | 0.711 | 0.009 | 1.299 | 8.7636 |
| M | 8342 2435 1 | 17 32 49.2 | -46 15 20 | 8.130 | 9.274 | 2.695 | 0.026 | 1.395 | 0.015 | 1.300 | 4.2359 |
| L | 5099 450 1 | 18 29 52.31 | -01 49 03.5 | 8.037 | 9.113 | 8.087 | 0.022 | 6.784 | 0.013 | 1.303 | 53.2422 |
| M | 8166 221 1 | 09 18 51.7 | -45 00 29 | 7.199 | 8.310 | 0.924 | 0.017 | -0.381 | 0.010 | 1.305 | 2.8965 |
| M | 7768 1742 1 | 12 14 51.8 | -44 06 42 | 7.550 | 8.658 | 1.369 | 0.016 | 0.063 | 0.010 | 1.306 | 3.0043 |
| M | 7576 1290 1 | 03 59 20.3 | -43 55 03 | 6.836 | 7.936 | 2.035 | 0.015 | 0.728 | 0.010 | 1.307 | 5.7121 |
| M | 7626 1963 1 | 06 37 32.7 | -44 21 27 | 9.053 | 10.165 | 2.188 | 0.035 | 0.870 | 0.018 | 1.318 | 2.1935 |
| M | 7540 746 1 | 01 16 52.0 | -44 02 09 | 9.533 | 10.599 | 2.950 | 0.034 | 1.631 | 0.017 | 1.319 | 2.5622 |
| M | 9158 876 1 | 03 36 12.2 | -73 58 26 | 7.617 | 8.757 | 2.179 | 0.017 | 0.858 | 0.011 | 1.321 | 4.1910 |
| L | 4801 1171 1 | 06 52 46.59 | -00 36 33.7 | 9.180 | 10.323 | 2.469 | 0.040 | 1.147 | 0.020 | 1.322 | 2.2901 |
| M | 8067 761 1 | 03 52 38.0 | -45 34 13 | 8.691 | 9.766 | 2.630 | 0.027 | 1.307 | 0.014 | 1.323 | 3.2231 |
| M | 8447 1615 1 | 22 53 15.5 | -45 08 52 | 6.849 | 7.965 | 2.344 | 0.016 | 1.021 | 0.010 | 1.323 | 6.4576 |
| M | 9138 1875 1 | 01 01 33.1 | -71 32 58 | 7.753 | 8.872 | 2.890 | 0.017 | 1.564 | 0.011 | 1.326 | 5.4649 |
| M | 7639 1799 1 | 06 47 43.7 | -44 43 09 | 7.784 | 8.918 | 2.039 | 0.018 | 0.708 | 0.011 | 1.331 | 3.6335 |
| M | 8003 281 1 | 22 39 27.6 | -43 16 21 | 8.323 | 9.500 | 2.577 | 0.022 | 1.243 | 0.013 | 1.334 | 3.5731 |
| M | 7831 528 1 | 14 46 57.6 | -44 45 01 | 9.070 | 10.275 | 2.428 | 0.036 | 1.093 | 0.019 | 1.335 | 2.4152 |
| M | 7576 1474 1 | 03 49 43.6 | -43 04 57 | 9.066 | 10.203 | 2.908 | 0.034 | 1.565 | 0.017 | 1.343 | 2.9842 |
| M | 7895 688 1 | 17 23 13.0 | -44 42 58 | 7.880 | 9.072 | 2.520 | 0.022 | 1.169 | 0.012 | 1.351 | 4.2635 |
| M | 7959 1868 1 | 20 06 11.0 | -44 20 30 | 7.867 | 8.992 | 2.565 | 0.019 | 1.213 | 0.011 | 1.352 | 4.3596 |
| M | 7703 848 1 | 09 29 06.8 | -44 48 04 | 8.506 | 9.642 | 2.140 | 0.024 | 0.782 | 0.013 | 1.358 | 2.7399 |
| M | 7547 1136 1 | 01 24 41.9 | -44 31 43 | 6.265 | 7.409 | 0.889 | 0.015 | -0.470 | 0.010 | 1.359 | 4.2741 |
| M | 8104 790 1 | 06 35 34.8 | -45 04 19 | 8.430 | 9.606 | 2.511 | 0.025 | 1.151 | 0.013 | 1.360 | 3.3479 |
| M | 8010 1510 1 | 22 42 26.1 | -44 05 19 | 9.361 | 10.451 | 2.330 | 0.045 | 0.968 | 0.020 | 1.362 | 1.9931 |
| M | 8103 2002 1 | 06 33 27.9 | -45 03 49 | 8.678 | 9.816 | 2.460 | 0.028 | 1.094 | 0.015 | 1.366 | 2.8870 |
| M | 9320 529 1 | 20 27 33.6 | -74 17 06 | 7.958 | 9.140 | 2.275 | 0.018 | 0.904 | 0.011 | 1.371 | 3.6582 |
| M | 7768 1778 | 12 12 27.2 | -43 59 58 | 7.915 | 9.081 | 1.581 | 0.018 | 0.204 | 0.011 | 1.377 | 2.7156 |
| M | 8040 786 1 | 01 38 03.7 | -46 05 02 | 6.975 | 8.139 | 2.426 | 0.015 | 1.045 | 0.010 | 1.381 | 6.1797 |
| M | 8279 584 1 | 14 43 32.3 | -46 25 37 | 8.361 | 9.559 | 2.776 | 0.020 | 1.392 | 0.011 | 1.384 | 3.8036 |
| M | 7548 766 1 | 01 36 41.5 | -44 48 20 | 8.567 | 9.744 | 1.914 | 0.025 | 0.527 | 0.013 | 1.387 | 2.3260 |
| L | 586 333 1 | 23 43 14.42 | 01 06 47.0 | 8.857 | 10.025 | 2.042 | 0.039 | 0.653 | 0.020 | 1.389 | 2.1426 |
| M | 8279 3079 1 | 14 38 13.5 | -45 52 19 | 6.834 | 8.010 | 2.758 | 0.016 | 1.367 | 0.010 | 1.391 | 7.6251 |
| M | 8342 1146 1 | 17 34 46.4 | -45 07 11 | 8.764 | 9.924 | 2.591 | 0.039 | 1.186 | 0.019 | 1.405 | 2.8682 |
| M | 9239 2066 1 | 12 03 44.5 | -74 12 51 | 6.444 | 7.666 | 0.713 | 0.016 | -0.694 | 0.010 | 1.407 | 3.5028 |
| M | 7537 404 1 | 01 16 12.0 | -42 00 35 | 6.580 | 7.784 | 2.183 | 0.015 | 0.765 | 0.010 | 1.418 | 6.4622 |
| M | 7959 1246 1 | 20 09 26.9 | -43 55 02 | 7.846 | 9.034 | 2.482 | 0.020 | 1.058 | 0.012 | 1.424 | 4.1776 |
| M | 8405 1805 1 | 20 19 36.9 | -46 25 41 | 8.717 | 9.894 | 8.861 | 0.034 | 7.432 | 0.017 | 1.429 | 51.1449 |
| M | 9363 459 1 | 03 38 39.6 | -75 45 46 | 7.674 | 8.866 | 3.041 | 0.018 | 1.606 | 0.011 | 1.435 | 5.7561 |
| M | 7576 1611 1 | 03 56 15.5 | -43 06 57 | 7.541 | 8.777 | 2.122 | 0.017 | 0.665 | 0.011 | 1.457 | 3.9730 |
| M | 8391 1112 1 | 20 02 36.2 | -45 11 46 | 6.567 | 7.791 | 1.404 | 0.016 | -0.054 | 0.010 | 1.458 | 4.4783 |
| L | 266 586 1 | 11 13 13.42 | 04 28 56.7 | 8.702 | 9.880 | 8.994 | 0.028 | 7.529 | 0.015 | 1.465 | 54.4599 |
| M | 8067 2458 1 | 03 53 33.3 | -46 53 37 | 5.932 | 7.167 | 2.417 | 0.015 | 0.948 | 0.009 | 1.469 | 9.4947 |
| M | 7895 1039 1 | 17 25 40.9 | -44 46 46 | 7.645 | 8.893 | 0.091 | 0.018 | -1.382 | 0.011 | 1.473 | 1.4767 |

| K, L, M | TYC | RA-2000 | Dec-2000 | V | B | bt | σBT | vt | σVT | χ | Plx |
|---|---|---|---|---|---|---|---|---|---|---|---|
| M | 8033 1022 1 | 01 25 30.6 | -47 13 38 | 9.356 | 10.637 | 1.692 | 0.040 | 0.211 | 0.018 | 1.481 | 1.4131 |
| L | 558 1634 1 | 22 06 01.54 | 01 22 47.9 | 8.933 | 10.120 | 2.013 | 0.046 | 0.530 | 0.022 | 1.483 | 1.9678 |
| M | 8168 1647 1 | 09 38 16.8 | -45 34 55 | 8.894 | 10.155 | 2.652 | 0.031 | 1.144 | 0.015 | 1.508 | 2.6791 |
| M | 8007 1434 1 | 22 47 09.1 | -41 41 31 | 6.834 | 8.114 | 1.366 | 0.016 | -0.157 | 0.010 | 1.523 | 3.7442 |
| M | 7576 2120 1 | 03 52 39.8 | -44 56 49 | 7.593 | 8.898 | 1.307 | 0.018 | -0.221 | 0.011 | 1.528 | 2.5623 |
| M | 8216 2577 1 | 11 51 08.7 | -45 10 24 | 4.466 | 5.767 | 0.486 | 0.014 | -1.044 | 0.009 | 1.530 | 7.4043 |
| M | 9412 2399 1 | 12 07 49.9 | -75 22 01 | 5.161 | 6.458 | 1.041 | 0.014 | -0.491 | 0.009 | 1.532 | 6.9109 |
| L | 5048 699 1 | 16 37 21.17 | -00 24 48.6 | 7.964 | 9.267 | 3.120 | 0.024 | 1.587 | 0.013 | 1.533 | 4.9624 |
| M | 7640 1263 1 | 06 51 26.1 | -44 27 27 | 8.170 | 9.475 | 0.974 | 0.022 | -0.560 | 0.012 | 1.534 | 1.6819 |
| M | 7754 1039 1 | 11 54 50.0 | -43 36 39 | 7.431 | 8.711 | 1.700 | 0.018 | 0.166 | 0.010 | 1.534 | 3.3091 |
| M | 8892 1165 1 | 05 46 01.2 | -61 13 47 | 7.405 | 8.691 | 1.347 | 0.017 | -0.188 | 0.011 | 1.535 | 2.8339 |
| M | 8392 1323 1 | 20 11 55.4 | -45 22 20 | 8.892 | 10.146 | 1.772 | 0.037 | 0.233 | 0.016 | 1.539 | 1.7561 |
| M | 9318 1152 1 | 19 43 22.8 | -74 46 09 | 7.588 | 8.905 | 1.615 | 0.017 | 0.061 | 0.010 | 1.554 | 2.9192 |
| M | 8003 365 1 | 22 37 19.3 | -44 39 55 | 7.917 | 9.169 | 1.160 | 0.021 | -0.397 | 0.012 | 1.557 | 2.0202 |
| M | 8447 1617 1 | 22 45 40.8 | -46 32 50 | 5.507 | 6.822 | 0.606 | 0.015 | -0.960 | 0.009 | 1.566 | 4.7597 |
| L | 4865 508 1 | 08 53 14.41 | -00 43 30.3 | 9.148 | 10.422 | 2.802 | 0.050 | 1.230 | 0.021 | 1.572 | 2.4754 |
| M | 9141 7568 1 | 01 00 53.7 | -72 41 51 | 7.692 | 9.030 | 1.559 | 0.018 | -0.037 | 0.011 | 1.596 | 2.6461 |
| M | 8033 281 1 | 01 18 01.6 | -45 21 03 | 8.946 | 10.248 | 1.099 | 0.030 | -0.497 | 0.014 | 1.596 | 1.2158 |
| L | 4966 1220 1 | 12 35 52.60 | -00 57 53.6 | 8.798 | 10.161 | 1.974 | 0.042 | 0.374 | 0.018 | 1.600 | 1.9424 |
| M | 8068 334 1 | 04 05 15.5 | -47 13 20 | 8.765 | 10.134 | 1.624 | 0.033 | -0.004 | 0.014 | 1.628 | 1.6478 |
| L | 447 476 1 | 18 41 26.68 | 00 33 51.8 | 7.474 | 8.923 | 1.974 | 0.018 | 0.274 | 0.010 | 1.700 | 3.3845 |
| M | 9315 718 1 | 20 18 27.0 | -72 58 46 | 6.560 | 7.970 | 1.430 | 0.015 | -0.275 | 0.010 | 1.705 | 3.9923 |
| M | 7547 1078 1 | 01 22 02.2 | -44 25 30 | 9.464 | 10.892 | 0.626 | 0.049 | -1.083 | 0.019 | 1.709 | 0.7242 |
| M | 8068 1578 1 | 04 05 40.8 | -46 48 07 | 8.501 | 9.929 | 1.142 | 0.029 | -0.569 | 0.014 | 1.711 | 1.4293 |
| M | 8447 1055 1 | 22 52 26.4 | -45 59 29 | 7.249 | 8.697 | 1.395 | 0.019 | -0.332 | 0.011 | 1.727 | 2.8109 |
| M | 7959 1673 1 | 20 06 55.2 | -44 18 11 | 8.104 | 9.520 | 1.509 | 0.023 | -0.225 | 0.012 | 1.734 | 2.0078 |
| M | 7896 351 1 | 17 36 41.6 | -44 52 45 | 7.383 | 8.815 | 2.035 | 0.022 | 0.298 | 0.012 | 1.737 | 3.5738 |
| M | 7547 781 1 | 01 32 44.0 | -44 44 21 | 8.323 | 9.775 | 1.934 | 0.026 | 0.188 | 0.012 | 1.746 | 2.2113 |
| M | 8343 1749 1 | 17 39 57.7 | -45 08 22 | 8.666 | 10.115 | 1.916 | 0.056 | 0.170 | 0.020 | 1.746 | 1.8188 |
| M | 7936 1985 1 | 06 46 46.5 | -44 58 27 | 7.978 | 9.465 | 0.446 | 0.024 | -1.305 | 0.012 | 1.751 | 1.2786 |
| M | 8280 955 1 | 14 59 18.6 | -45 26 35 | 8.821 | 10.373 | 0.194 | 0.045 | -1.562 | 0.017 | 1.756 | 0.7611 |
| M | 7896 3400 1 | 17 38 30.5 | -44 52 32 | 7.931 | 9.449 | 1.289 | 0.032 | -0.479 | 0.014 | 1.768 | 1.9137 |
| M | 8007 31 1 | 22 43 17.7 | -41 32 36 | 8.253 | 9.805 | 1.491 | 0.027 | -0.288 | 0.013 | 1.779 | 1.7994 |
| L | 4896 1281 1 | 09 56 39.11 | -00 27 41.2 | 7.835 | 9.320 | 1.264 | 0.029 | -0.519 | 0.014 | 1.783 | 1.9485 |
| M | 8033 836 1 | 01 27 29.3 | -45 50 46 | 6.961 | 8.455 | 1.003 | 0.016 | -0.782 | 0.010 | 1.785 | 2.6163 |
| M | 9412 2483 1 | 12 04 46.5 | -76 31 09 | 5.025 | 6.508 | 1.111 | 0.015 | -0.678 | 0.009 | 1.789 | 6.6724 |
| M | 485 197 1 | 20 01 04.1 | -44 23 07 | 8.880 | 10.406 | 1.809 | 0.044 | 0.018 | 0.016 | 1.791 | 1.5651 |
| M | 7583 1577 1 | 04 12 31.6 | -44 22 06 | 6.705 | 8.187 | 0.491 | 0.016 | -1.301 | 0.010 | 1.792 | 2.3250 |
| L | 5278 1473 1 | 01 57 31.92 | -07 32 09.7 | 8.192 | 9.704 | 1.449 | 0.034 | -0.346 | 0.015 | 1.795 | 1.7951 |
| M | 7540 576 1 | 01 18 44.2 | -43 19 58 | 6.766 | 8.249 | 1.098 | 0.016 | -0.709 | 0.010 | 1.807 | 2.9586 |
| L | 4966 1089 1 | 12 35 59.53 | -00 34 39.6 | 8.309 | 9.830 | 1.612 | 0.034 | -0.196 | 0.010 | 1.808 | 1.8444 |
| M | 8104 240 1 | 06 44 35.7 | -46 46 25 | 7.687 | 9.201 | 1.108 | 0.023 | -0.713 | 0.012 | 1.821 | 1.9249 |

Conversion of Tycho-2 to Johnson-Cousins magnitudes in the Gaia era

| K, L, M | TYC | RA-2000 | Dec-2000 | V | B | bt | σBT | vt | σVT | χ | Plx |
|---|---|---|---|---|---|---|---|---|---|---|---|
| M | 8392 934 1 | 20 05 07.9 | -46 05 49 | 6.954 | 8.463 | 1.369 | 0.018 | -0.465 | 0.010 | 1.834 | 3.0156 |
| M | 8216 235 1 | 11 56 38.9 | -45 32 48 | 8.534 | 10.063 | 1.454 | 0.030 | -0.389 | 0.012 | 1.843 | 1.5172 |
| M | 7547 1134 1 | 01 28 21.9 | -43 19 06 | 3.400 | 4.971 | 1.315 | 0.014 | -0.568 | 0.009 | 1.883 | 14.5824 |
| M | 7626 470 1 | 06 39 26.8 | -43 25 38 | 7.172 | 8.722 | 2.061 | 0.019 | 0.176 | 0.010 | 1.885 | 3.6509 |
| M | 7832 1936 1 | 14 55 48.4 | -44 41 48 | 7.939 | 9.489 | 1.818 | 0.023 | -0.070 | 0.011 | 1.888 | 2.2991 |
| M | 8104 1857 1 | 06 36 50.9 | -46 01 33 | 8.630 | 10.213 | 0.843 | 0.037 | -1.053 | 0.014 | 1.896 | 1.0851 |
| M | 8010 1347 1 | 22 51 39.8 | -44 59 50 | 8.704 | 10.258 | 1.712 | 0.049 | -0.216 | 0.017 | 1.928 | 1.5275 |
| M | 7583 905 1 | 04 04 47.7 | -44 40 07 | 8.022 | 9.620 | 1.061 | 0.028 | -0.875 | 0.013 | 1.936 | 1.5324 |
| M | 9364 1159 1 | 04 14 30.9 | -75 48 29 | 7.239 | 8.884 | 1.650 | 0.019 | -0.290 | 0.011 | 1.940 | 2.8416 |
| L | 511 1907 1 | 20 38 16.46 | 01 00 59.4 | 7.885 | 9.526 | 1.021 | 0.027 | -0.922 | 0.012 | 1.943 | 1.5927 |
| L | 5176 1533 1 | 20 37 07.63 | -00 31 02.6 | 8.711 | 10.373 | 1.103 | 0.047 | -0.928 | 0.015 | 2.031 | 1.0857 |
| L | 479 1337 1 | 19 38 47.14 | 00 36 20.3 | 8.292 | 10.008 | 2.402 | 0.032 | 0.369 | 0.013 | 2.033 | 2.3707 |